%% file: manuscript.tex
\documentclass[fleqn,usenatbib]{mnras}
\pdfoutput=1

\usepackage[T1]{fontenc}
\usepackage{ae,aecompl}
\usepackage{tablefootnote}

\usepackage{graphicx}	% Including figure files
\usepackage{amsmath}	% Advanced maths commands
\usepackage{amssymb}	% Extra maths symbols

\usepackage{newtxtext,newtxmath}
\usepackage{lineno,multirow}
\title[Evolution of M Dwarf Planet Radii]{The Radius Distribution of M dwarf-hosted Planets and its Evolution}
\author[Gaidos et al.]{Eric Gaidos$^{1,2}$, 
\thanks{E-mail: gaidos@hawaii.edu.}
Aleezah Ali$^{3}$,
Adam L. Kraus$^{4}$,
Jason F. Rowe$^{5}$
\\
$^{1}$Department of Earth Sciences, University of Hawai'i at M\={a}noa, 1680 East-West Rd, Honolulu, HI 96822, USA\\
$^{2}$Institute for Astrophysics, University of Vienna, T\"{u}rkenschanzstrasse 17, 1180 Vienna, Austria\\
$^{3}$Institute for Astronomy, University of Hawai'i at M\={a}noa, 2680 Woodlawn Drive, Honolulu, HI 96822, USA\\
$^{4}$Department of Astronomy, University of Texas at Austin, 2515 Speedway Stop C1400, Austin, TX 78712, USA\\
$^{5}$Department of Physics \& Astronomy, Bishop's University, 2600 College Street
Sherbrooke, QC, Canada, J1M 1Z7 
}

\date{}

\pubyear{}

\hypersetup{draft}
\begin{document}
\include{newcommands}
\label{firstpage}
\pagerange{\pageref{firstpage}--\pageref{lastpage}}
\maketitle

% Abstract of the paper  (200 words for letters)
\begin{abstract}
M dwarf stars are the most promising hosts for detection and characterization of small and potentially habitable planets, and provide leverage relative to solar-type stars to test models of planet formation and evolution.  Using \gaia\ astrometry, adaptive optics imaging, and calibrated gyrochronologic relations to estimate stellar properties and filter binaries we refined the radii of 117 \kepler{} Objects of Interest (confirmed or candidate planets) transiting 74 single late K- and early M-type stars, and assigned stellar rotation-based ages to 113 of these.  We constructed the radius distribution of 115 small ($<$4\rearth) planets and assessed its evolution.  As for solar-type stars, the inferred distribution contains distinct populations of ``super-Earths" (at $\sim$1.3\rearth) and ``sub-Neptunes" (at $\sim$2.2 \rearth) separated by a gap or ``valley" at $\approx$1.7\rearth that has a period dependence that is significantly weaker (power law index of -0.03$^{+0.01}_{-0.03}$) than for solar-type stars.  Sub-Neptunes are largely absent at short periods ($<$2 days) and high irradiance, a feature analogous to the ``Neptune desert" observed around solar-type stars.  The relative number of sub-Neptunes to super-Earths declines between the younger and older halves of the sample (median age 3.86 Gyr), although the formal significance is low ($p = 0.08$) because of the small sample size.  The decline in sub-Neptunes appears to be more pronounced on wider orbits and low stellar irradiance.  This is not due to detection bias and suggests a role for \water{} as steam in inflating the radii of sub-Neptunes and/or regulating the escape of H/He from them.
\end{abstract}


%%%%%%%%%%%%%%%%%%%%%%%%%%%%%%%%%%%%%%%%%%%%%%%%%%

%%%%%%%%%%%%%%%%% BODY OF PAPER %%%%%%%%%%%%%%%%%%

\section{Introduction}
\label{sec:intro}

The launch of the \kepler\ mission in 2009 \citep{Borucki2010} ushered in a revolution in the study of planetary systems, the occurrence of rocky, Earth-like planets, and the potential for habitable and life-hosting planets in the Galaxy.  \kepler\ revealed a population of small (Earth- to Neptune-size) planets on close-in orbits around solar-type stars \citep[e.g.,][]{Burke2015,Hsu2019,Bryson2021}.  Analyses showed that the occurrence of these planets increases with decreasing stellar mass or later spectral type \citep{Dressing2015,Mulders2015,Gaidos2016,Hardegree-Ullman2019,Ment2023}.  This trend is important to the search for Earth-like planets since the low luminosities and compact ``habitable zones" of M dwarfs mean that they have the potential to host many rocky planets with temperate climates; the elevated space density of M dwarfs relative to their solar-mass counterparts \citep{Smart2021,Ravinet2021} means that many host stars are nearby and can thus be prime targets for follow-up observations to detect and characterize atmospheres, e.g., by \jwst\ \citep{Greene2023,Zieba2023},

However, M dwarfs differ from their solar-mass counterparts in ways that could impact the properties and evolution of their planets, including an extended pre-main sequence phase \citep{Luger2015}, longer rotational spin-down times \citep[e.g.,][]{Johnstone2021}, and elevated X-ray and extreme ultraviolet emission relative to bolometric emission \citep[e.g.,][]{France2016}.  There may also be differences in their protoplanetary disks, i.e. their masses \citep{Gaidos2017b}, lifetimes \citep{Silverberg2020}, and chemistry \citep{Tabone2023}.  To assess potential variation in planetary evolution, precise measurement of the properties of a large number of planets in systems with robust ages that span a wide age range is needed.  Currently, the confirmed and candidate transiting planets among the \kepler\ Objects of Interest (KOIs) provide the largest, most uniform, and most precisely measured set  of planet properties such as radii and periods \citep[e.g.,][]{Thompson2018}.  

The radius distribution of sub-Jovian planets around solar-type stars detected by the \kepler{} mission contains two distinct peaks, representing populations of ``super-Earths"- and ``sub-Neptunes" separated by a gap or ``valley" at $\sim$1.7\rearth\ \citep{Owen2013,Fulton2017,Berger2020}.    Complementary measurements of planet mass (and hence density) indicate that sub-Neptunes and super-Earth have rocky cores with and without envelopes of volatile, low molecular-weight constituents \citep{Weiss2014,Rogers2015}.  These envelopes could consist of hydrogen and helium captured from the primordial gas disk during planet formation \citep[e.g.,][]{William2019}, and/or H$_2$O and other volatiles acquired as ices \citep{Turbet2020,Luque2022}.  Unfortunately, mass measurements of small planets are observationally expensive, often not precise, and the mass-radius relations are at least partly degenerate with composition \citep[e.g.,][]{Rogers2010}.

Pursuing an alternative approach, numerous statistical studies have made inferences about the nature of these planets and their manner of their evolution by correlating variation in the radius distribution with parameters such as orbital period (semi-major axis) and host star mass \citep[][and references therein]{ZhuW2021}.  The former sets the stellar irradiation of the planet and the latter -- with M dwarfs at one extreme-- governs the timescale over which that irradiation changes \citep{Barnes2009,Johnstone2021}.  Due to their faintness, limited inclusion in the \kepler{} survey, and more poorly understood properties of M dwarf stars, studies of their planets' radius distribution have lagged those of solar-type stars \citep[e.g.,][]{Dressing2015,Gaidos2016,Hirano2018,Kanodia2019}.  

The question of planetary evolution has also been addressed directly by looking for variation with host star age.  Among FGK stars, the number of sub-Neptunes (with envelopes) relative to super-Earths (without envelopes) planets is found to decrease with age \citep{Berger2020,Sandoval2021,David2021,ChenD-C2022}, consistent with cooling and contraction of the envelopes and/or escape of light elements to space \citep[e.g.,][]{Howe2015}.  The mean size of sub-Neptunes may also decrease with time \citep{ChenD-C2022} -- but see \citet{Petigura2022}. 

Reconstructing the planet radius distribution of M dwarf planets and its evolution requires radii and ages that are more precise and accurate than have previously been available.  Transiting planet radius depends on host star radius, and calculations of the radius distribution that account for detection efficiency require stellar density (both mass and radius).  New empirical calibrations for M dwarfs permit estimation of mass and radius based on the precise absolute luminosities made possible by \gaia\ parallaxes \citep{Mann2015,Mann2019}.  Moreover, both \gaia\ astrometry and ground-based adaptive optics (AO) imaging allow more thorough vetting for stellar companions, which, if not properly accounted for, can affect a radius estimate.  Binary stars also appear to have a planet population distinct from single stars \citep{Kraus2016,Sullivan2023}.

For field M dwarfs, rotation (gyrochronology) is the most promising method to assign ages compared to other techniques (i.e., isochrone-fitting, asteroseismology, depletion of lithium, or kinematics), but in the absence of accurate, first-principles spin-down models, this approach requires empirical calibration.   \citet{Dungee2022} constructed a rotation sequence (vs. effective temperature \teff) for late K-type and early-type M dwarfs in the 4 Gyr-old M67 cluster, extended calibrated age-dating of \teff$\approx$3200--4200K stars to nearly half the age of the Galactic disk.  Moreover, a comparison with younger clusters suggests these stars are spinning down according to a simple power-law prescription \citep{Skumanich1972} by ages of a few Gyr, justifying an extrapolation to older ages.  \citet{Gaidos2023b} constructed \teff-dependent age-rotation relations using rotation sequences and assigned ages to the cool host stars of known exoplanets, including \kepler\-confirmed and candidate planets.

In this work we revise the radius distribution of \kepler{} Objects of Interest (KOIs, i.e., confirmed, validated or candidate planets) around single late K- and early M-type dwarf stars and examine its variation with age, taking advantage of precise \gaia\ parallaxes for luminosity calculations and revised mass/radius/luminosity relations based on nearby stars; identification and screening of binary systems using astrometry and AO imaging; and assignment of robust rotation-based ages based on the \citet{Gaidos2023b} gyrochronology.  We compare the M dwarf planet radius distribution to that derived for solar-type stars and interpret it and its change with time to different scenarios for planet composition and evolution.  

\section{Analysis}
\label{sec:analysis}

We matched the Kepler Input Catalog \citep[KIC,][]{Brown2011} with \gaia\ Data Release 3 (DR3) sources \citep{Gaia2023} using an angular separation criterion of $<$0".71.  Absolute $K_s$-band (2.2 \micron) magnitudes ($M_K$) were computed using photometry from the 2MASS Point Source Catalog \citep{Skrutskie2006,Cutri2013} and parallaxes from \gaia\ Data Release 3 \citep{Gaia2016,Gaia2023}. We restricted our analysis to stars observed by \kepler\ during Quarters 1-17 with $M_K > 4.5$ \citep[inclusive of spectral types later than K5,][]{Pecaut2013}, and that appear in the \citet{Santos2019} catalog of late K and M-type dwarfs with rotation signals from \kepler\ lightcurves (about 59\% of the 24,162 stars analyzed).  We further limited the sample to stars with \teff\ values in \citet{Santos2019} between 3200 and 4200K, the valid limits of the \citet{Gaidos2023b} gyrochronology to be applied to these stars , yielding a final sample of 5433 target stars that \kepler\ observed with the relevant host star parameters.   We identified KOIs around these stars from the DR25 catalog, which is based on data from Quarters 1--17 \citep{Thompson2018}.  

We excluded KOIs that have false positive (FP) dispositions in the DR25 Supplemental Catalog, which includes a few planets (candidates) erroneously categorized as false positives in DR25.\footnote{https://exoplanetarchive.ipac.caltech.edu/docs/PurposeOfKOITable.html\#q1-q17\_sup\_dr25}.  We also excluded those with a disposition of ``candidate" \emph{and} a modified astrophysical FP probability $>0.01$ calculated with {\tt vespa} \citep{Morton2015}.  This threshold was chosen based on the distributions of FP probability; above $p>0.01$, FPs outnumber confirmed or validated planets.   The modified probability is the sum of the probabilities of the undiluted eclipsing binary (EB) and hierarchical EB scenarios, plus the background EB scenario if our AO imaging cannot exclude this.  The latter entails that imaging was obtained, that there are no additional sources detected within 2" of the KOI \citep[and thus falling on the same \kepler{} pixel and inducing minimal centroid motion,][]{Adams2012} and that the contrast limit of the imaging is sufficient to detect an EB sufficiently bright to produce the observed signal.  For this we adopt a maximum eclipse depth of 50\% (zero impact parameter) and a \kepler{}-$K_s$ reddening of 0.24 for background sources.  The latter is based on a mean total reddening $E(B-V) = 0.12$ for the \kepler{} field \citep{Schlafly2011} and the extinction coefficients of \citet{Yuan2013}, substituting Sloan $r$ for the \kepler{} bandpass.  Hereafter we refer to all KOIs that pass these tests as ``planets".

Host star masses were calculated from an empirical mass-luminosity relationship \citep[Eqn. 4 in ][]{Mann2019}.  Radii were calculated from $M_K$ using Eqns. 4 or 5 in \citet{Mann2015} for stars with or without estimates of [Fe/H] retrieved from the Exoplanet Archive.\footnote{These [Fe/H] estimates are typically from published analyses, to be distinguished from the metallicity estimates included in the original KIC.}  Errors were estimated using Monte Carlo calculations assuming normally-distributed errors in $K_s$-band magnitudes and parallaxes.  Finally, we identified those KOIs with gyrochronologic age assignments in \citet{Gaidos2023b}.     
\subsection{Filtering of binary stars}
\label{sec:binaries}

Stars in binary or multiple systems with separations $\lesssim$50 au are known to host a planet population different from that around single stars \citep{Kraus2016}.  Furthermore, binary stars with separations ($<$100~au) have different rotational histories than their single counterparts \citep[e.g.,][]{Messina2017,Stauffer2018b}, meaning that a rotation-age relation developed for single stars cannot be applied.   90\% of M dwarfs in the \kepler\ field are more distant than 135 pc, thus these physical separations correspond to angular separations $\lesssim$1".  Companions or any nearby sources unresolved or partially resolved by \kepler\ (which had 4" pixels and a 3.1-7.5" PSF) can also dilute transit signals and make identification of the host star itself -- and thus planet properties -- ambiguous.  The $K_s$-band photometry used to estimate radius and mass will be inaccurate for binaries that are unresolved by the 2MASS survey (resolution of $\sim$3").  Thus we identified and excluded KOIs around binary stars with separations $<4$".   Binary/multiple systems or candidate systems were identified and excluded using (1) \gaia\ astrometry of resolved sources; (2) \gaia\ astrometric error relative to single-star astrometric solutions; (3) adaptive optics imaging; and (4) over-luminosity relative to a single-star locus in a color-absolute magnitude diagram. 

\textit{\gaia\ resolved companions:} We identified companions resolved by the \gaia{} satellite \citep[$\gtrsim$1 arcsec,][]{Ziegler2018} based on similarity of parallax (distance) and proper motion.  Rather than fixed criteria for similarity \citep[e.g.,][]{El-Badry2021}, we adopted a Bayesian approach.  We identified candidate companions with cone searches of the DR3 catalog with a radius of 74" (corresponding to about $10^4$ au at the median distance of \kepler\ M dwarfs) around each \kepler{} M dwarf.  We computed the probability of companionship as $p_{\rm binary} = P_c/(P_c + P_b)$, where $P_c$ and $P_b$ are the likelihoods that the source is a companion or background star, respectively.   We evaluated these assuming normally-distributed errors using:
\begin{equation}
\label{eq:Pc}
    P_{c} = \frac{\Delta \mu}{\sigma_{\Delta \pi} \left(\sigma_{\Delta\mu}^2+\mu_{\rm orb}^2\right)} \exp \left[-\frac{\Delta \mu^2}{2\left(\sigma_{\Delta \mu}^2 + \mu_{\rm orb}^2\right)}
    -\frac{\Delta \pi^2}{2\sigma_{\Delta \pi}^2}\right],
\end{equation}
and
\begin{equation}
\label{eq:Pb}
    P_{b} = \frac{1}{N-1}\sum_{i=1}^{S} \frac{\Delta \mu_i}{\sigma_{\Delta \pi} \left(\sigma_{\Delta\mu}^2+\mu_{\rm orb}^2\right)} \exp \left[-\frac{\Delta \mu_i^2}{2\left(\sigma_{\Delta \mu}^2 + \mu_{\rm orb}^2\right)}
    -\frac{\Delta \pi_i^2}{2\sigma_{\Delta \pi}^2}\right],
\end{equation}
where $\Delta \mu$ and $\sigma_{\Delta \mu}$ are the absolute scalar difference in proper motion between the star and source and its standard error, $\Delta \pi$ and $\sigma_{\Delta \pi}$ are the difference in parallax and its standard error, and $\mu_{\rm orb}$ is a ``softening" term that accounts for the angular orbital motion of a binary.  This last term we set to 5 mas yr$^{-1}$ based on Eqn. 4 in \citet{El-Badry2018}.  Constant factors that divide out are omitted from Eqns. \ref{eq:Pc} and \ref{eq:Pb}.   The difference in the terms involving parallax and proper motion reflects the latter's two-dimensionality.  To improve the statistical accuracy by averaging over many stars, the summation for $P_b$ is performed over the $S$ stars in all the $N-1$ \emph{other} cone searches (a few thousand stars) that cannot be companions. We imposed no weighting with angular separation other than the generous search radius.  Based on the observed distribution of $P_{\rm binary}$ values relative to those calculated using stars from other fields, we identified a threshold of $P_{\rm binary} > 0.98$ to select companions.

\textit{Gaia astrometric error:} Binaries with separations $<1$" cannot be resolved by \gaia\ \citep{Ziegler2018b} but can often be identified based on the poor fit of \gaia\ astrometry to a single-star solution \citep{Belokurov2020}.  This is quantified as the Reduced Unit Weight Error \citep[RUWE,][]{Lindegren2018} and values significantly greater than unity are indicative of a multiple star system.  We selected a threshold of 1.4 based on \citet{Lindegren2021}.

\textit{AO imaging:} AO imaging of KOIs was obtained of all but 8 systems with the NIRC2 infrared imager and AO system on the Keck-2 telescope \citep{McLean2004}.  Observations used the narrow camera (10 mas pixel scale) and either the $K_{cont}$ or $K_p$ filters.  Images were processed, sources extracted, and astrometry performed using the pipeline described in \citet{Kraus2016}.  We selected those sources with separations $<$2" and contrast $\Delta K < 4$ as highly likely to be bound companions based on \citet{Kraus2016}.

\textit{Photometry:} Unresolved late K and M dwarf binaries can be identified by their over-luminosity relative to single stars on the main sequence since these stars evolve imperceptibly on the main sequence.  Metallicity is the sole confounding factor, and differences in [Fe/H] produce a dispersion in color-absolute magnitude relations.\footnote{Gravity does not vary significantly for a given \teff\ among late K- and M dwarf stars and extinction is negligible at the distance of most M dwarfs observed by \kepler.}  By experimenting with different color-magnitude combinations constructed from ground-based photometry, we found that a color-magnitude locus of \kepler\ M dwarfs constructed using PanSTARRS $g$-$y$ vs. $M_y$ \citep{Tonry2012} has the smallest dispersion and is thus the least sensitive to metallicity.  We identified stars that are more luminous than the best fit by more than two standard deviations as probable binaries. 

\subsection{Planet radius distribution}

The posterior distribution in radius of each planet was determined by multiplying each value from the Monte Carlo Markov Chain (MCMC) posterior chains for planet-to-star radius ratio from DR25 Q1-17 \citep{Hoffman2017} with a value from the stellar posterior distribution in radius calculated above, assuming Gaussian errors in magnitude and parallax.  To correct for detection efficiency, we normalized each distribution by the summed probability for detection of an equivalent planet around every star observed by \kepler\ that is a late K- or early M-type dwarf $M_K > 4.5$, 3200 $<$ \teff\ $<$ 4200K) and that appears in the \citet{Santos2019} catalog of stars evaluated for rotation periods. We used the {\tt KeplerPORTs} code \citep{Burke2017} to calculate the probability of detection (including the geometric transit probability).  The required window functions, one-sigma detection functions \citep{Burke2017b}, ``dataspans" (timespans of data), and duty cycles  were retrieved from the NASA Exoplanet Archive.    Non-linear (four-coefficient) limb-darkening coefficients were obtained from the DR25 Stellar Properties Catalog \citep{Mathur2017}.  To mimic the removal of binaries from the KOI list, we excluded all target stars with \gaia\ RUWE values $>$1.4.  Finally, KOIs were then matched with entries in the TIME Table of \citet{Gaidos2023b} to retrieve gyrochronologic age estimates.   

The summed probabilities over all stars represent the expected number of detections if each star in the observed sample had such a planet.  Each posterior distribution in radius was divided by this quantity and then summed to give the number of objects per unit radius per star.  As a metric of uncertainty due to radius error and finite counting statistics, we calculated 68\% confidence intervals by constructing 1000 Monte Carlo realizations using bootstrap sampling of the planet population with replacement.\footnote{While multi-planet systems will affect the counting statistics of total planet occurrence (which is not considered here), correlation between the radii of planets in the same system \citep[e.g.,][]{Weiss2018} would impact the significance of individual features.}  

\subsection{Validation}
\label{sec:validation}

\begin{figure}
\centering
\includegraphics[width=\columnwidth]{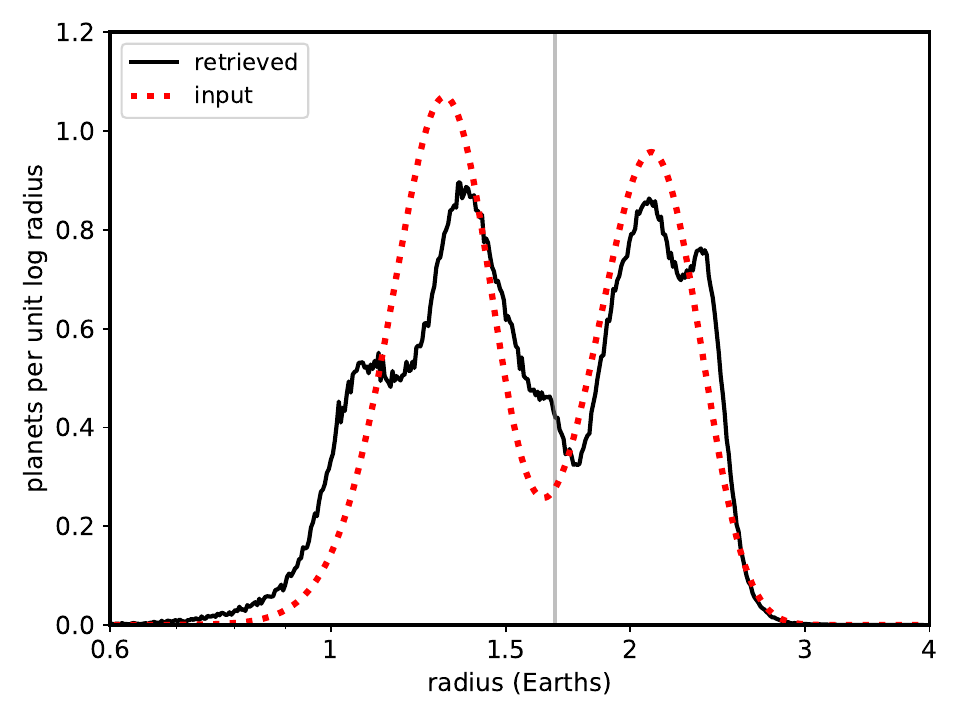}
\caption{Radius distribution of simulated planets (red dotted curve) and distribution retrieved from 115 simulated KOIs (solid black curve).  The vertical grey line at 1.68 \rearth{} is the approximate divide between super-Earths and sub-Neptunes found among actual KOIs.}
\label{fig:simulation}
\end{figure}

To validate our methodology for reconstructing the planet radius distribution we constructed a sample of synthetic KOIs the same size as the actual one (115 planets, see Sec. \ref{sec:results}), drawing radii from the sum of two Gaussian distributions representing super-Earths and sub-Neptunes.  The fractional populations in the two distributions were set to 0.4 and 0.6, the centroids were set to 1.3 and 2.1 \rearth, and the standard deviations were set to 0.15 and 0.25\rearth, respectively.  These values produces a reasonable radius distribution that roughly approximates the observed distribution but because this a test of the fidelity of recovery, the precise values do not matter.  Orbital periods were selected from a power-law distribution between 1 and 40 days that reproduces the observed distribution \citep{Gaidos2016}.  Synthetic planets were randomly assigned to the \kepler{} late K/early M population with measured rotation periods as defined above.  To calculate the planet-star radius ratio, the stellar radii were determined using the metallicity-dependent relation of \citet{Mann2015}, with metallicities drawn from the Gaussian distribution found among \kepler{} M dwarfs \citet{Gaidos2016}.  However, for added realism these artificial metallicities were not propagated to the analysis of synthetic KOIs since the metallicities of many KOIs are not well determined; instead, the metallicity-independent relation of \citet{Mann2015} was used.  The \emph{total} probability of detection (including the geometric factor) was calculated for each star-planet pair as described above and the pair accepted conditional to a random uniform draw below this probability.  In lieu of actual posterior distributions for the \emph{observed} radius ratio we adopted Gaussians with fractional standard deviation that are 1/SNR, where SNR is the predicted transit signal-to-noise \citep{Howard2012}:
\begin{equation}
\label{eqn:snr}
{\rm SNR} = \frac{\delta}{\sigma_{\rm CDPP}}\sqrt{\frac{N \tau}{2\,{\rm hr}}},
\end{equation}
where $\delta$ is the transit depth (approximately the radius ratio squared), $\sigma_{\rm CDPP}$ is the measured combined differential photometric precision of the star over 2 hours \citep[][, 2 hr is a typical transit duration]{Christiansen2012}, $N$ is the number of transits (equal to the dataspan times the duty cycle divided by the orbital period), and $\tau$ the transit duration (after averaging over a uniformly distributed impact parameter).  Host star-planet pairs were generated until the accepted number equaled the size of the actual sample  
We ran the synthetic KOIs through the same analysis pipeline as the actual KOIs to produced a simulated observed radius distribution.  This is plotted along with the input distribution in Fig. \ref{fig:simulation}.  The bimodal distribution is largely reproduced, but with a noticeable broadening of the super-Earth distribution, probably because of the absence of measured [Fe/H] "measurements" and the effect of noise on the radii of the smallest detectable planets.  The input ratio of super-Earths to Neptunes (divided by the vertical grey line) is 0.74 while the retrieved ratio is 0.78, an increase of 5\%.  The increase is due to the greater scatter of the more poorly determined super-Earths into the sub-Neptune range compared to vice-versa.  The location of the apparent valley is shifted to a slightly larger radius (from 1.64 to 1.75 \rearth) for the same reason.  

\section{Results}
\label{sec:results}

\subsection{Planet Radius-Period Distribution} 
\label{sec:radius-period}

We identified 117 confirmed/validated or candidate planets around 74 single late K and early M dwarfs with matches to \gaia\ DR3 sources and rotation periods in \citet{Santos2019}.  Thirteen have NASA Exoplanet Archive dispositions of "candidate" while the rest are "confirmed" (i.e., validated).  Orbital periods, radii and ages of KOIs are provided in Table \ref{tab:kois}. Of these, 114 (in 73 systems) have rotation-based ages:  nearly all are from \citet{Gaidos2023b} but two missing systems (KOIs 2542 and 2705) were assigned ages using the methods of \citet{Gaidos2023b}.  Four planets in three other systems (KOIs 430, 4957, and 6863) have rotation periods in \citet{Santos2019} that are shorter than the calibrated range in \citet{Gaidos2023b} and are not age-dated.   

\begin{figure}
    \centering
    \includegraphics[width=\columnwidth]{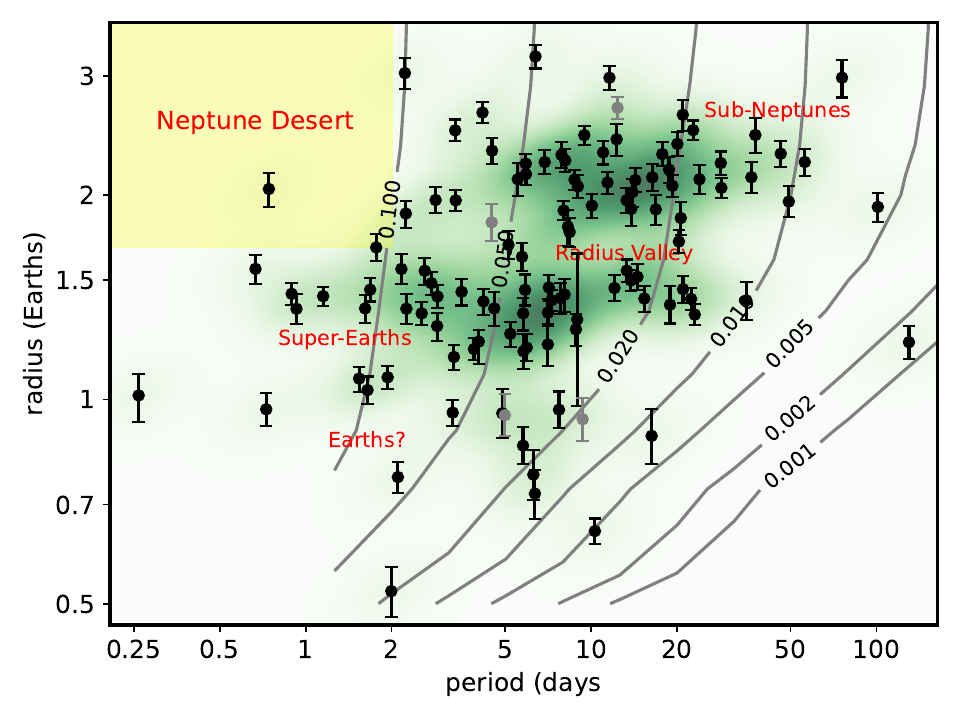}
    \caption{Orbital periods and radii of 115 transiting planets and planet candidates around late K and early M dwarfs detected by \kepler.  Black points represent those planets with rotation-based ages in \citet{Gaidos2023b}.  The shading is a Gaussian KDE with a bandwidth of 0.25 and major features of the distribution are labeled.  The grey contours demarcate the \kepler\ detection efficiency (including the geometric factor) for the DR25 catalog with late K and early M dwarf host stars of all ages, i.e. the fraction of planets detected if every star in the sample hosted a planet with a given radius and orbital period.}
    \label{fig:period-radius-kde}
\end{figure}

The two largest objects in the sample, excluded from further analysis, are the hot Jupiter KOI-254.01  \citep[Kepler-45b,][]{Johnson2012b}, and the $5.45\pm0.27$ \rearth\ KOI-2715.01 \citep[Kepler-1321b,][]{Morton2016} orbiting a late K dwarf (neither shown in Fig. \ref{fig:radius}).  The next largest planet is the $3.20 \pm 0.14$\rearth\  sub-Neptune KOI-253.01 \citep[Kepler-2000b,][]{Valizadegan2023}, highlighting how Neptune-size and larger planets are very rare around M dwarfs.  Figure \ref{fig:period-radius-kde} plots the radius vs. orbital period of the 115 small planets, along with a Gaussian kernel density estimator (KDE) to highlight structure.  Planets with assigned ages are plotted as black points; while those lacking ages are plotted in grey. 

We calculated the detection efficiency (i.e. fraction of detected planets with a given radius and period if distributed uniformly among the late K/early M type stars), in among 3200-4200K stars in \citet{Santos2019} using the procedures described in Sec. \ref{sec:analysis}.  This is represented as the grey contours in Fig. \ref{fig:period-radius-kde}.  As expected, detection efficiency is lowest for  smaller planets on wider orbits, which explains the absence of detections in the lower right corner of the plot.  There is a single object, KOI-571.05, which is in a region of particularly low detection efficiency ($<$0.002) and would normally warrant scrutiny as a potential false alarm, but this was validated as a member of the multi-planet Kepler-186 system by \citet{Quintana2014}.

As was shown for solar-type host stars \citep{Fulton2017} and subsequently for late K- and M-type dwarfs \citep{Hirano2018,Cloutier2020,VanEylen2021}, the planet distribution of this sample contains two distinct clusters at about 1.3\rearth\ (``super-Earths") and 2.2\rearth\ (``sub-Neptunes"), with an intervening gap or ``valley" at 1.6-1.7\rearth.  Around solar-type hosts there is a deficit of (sub-)Neptunes on short--period (or high--irradiance) orbits, the so-called ``Neptune desert" \citep{Szabo2011,Mazeh2016}.  There are no universally accepted boundaries for this desert but it is expected to depend on stellar mass \citep{McDonald2019}.  Adopting the conservative criterion of a period of $<$2 days (corresponding to irradiances $S\gtrsim$ 200\searth), the ``desert"  has a single unambiguous denizen in our sample: KOI--952.05 (Kepler--32f), a member of a multi-transiting planet system \citep{Swift2013}, has an orbital period of 0.74 days and radius of $2.04 \pm 0.11$\rearth, consistent with but more precise than that determined by \citet{Morton2016} and \citet{Berger2018b} but significantly larger than previous estimates \citep[e.g.,][]{Swift2013,Mann2013}.  There are five other Earth- and super-Earth-size Ultra Short Period (USP, $P<1$~d) planets in the sample.  The smallest object in the sample is the $0.52^{0.04}_{0.05}$\rearth{} candidate KOI--7793.01.  

\begin{figure}
    \centering
    \includegraphics[width=\columnwidth]{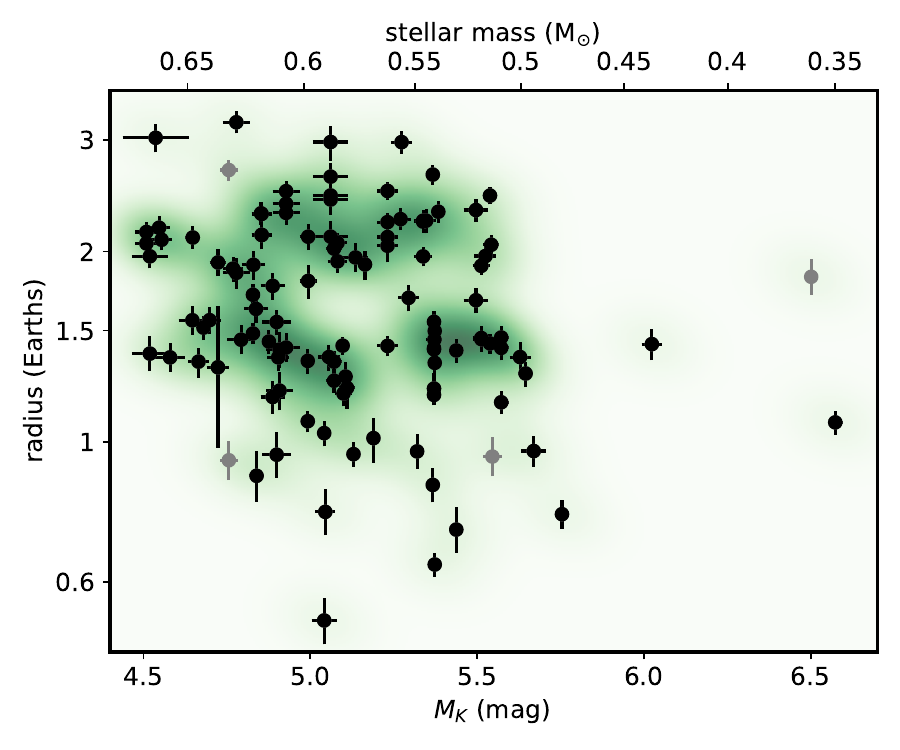}
    \caption{Planet radius vs. stellar luminosity (absolute $K_s$ magnitude) and stellar mass.  Black point indicate KOIs in systems with ages assigned by gyrochronology.   Multi-planet systems appear as vertical chains of points.  The shading is a Gaussian KDE with a bandwidth of 0.25.}
    \label{fig:rad_mk}
\end{figure}

\citet{Cloutier2020} found that the planet radius distribution hosted by stars with masses 0.08--0.42\msun\ (corresponding to spectral types M3-M8) differs from that of more massive stars in lacking a significant population of (sub)Neptunes.  \citet{Ment2023} also found a very pronounced dearth of sub-Neptunes around mid-to-late type M dwarfs.  Our sample does not substantially overlap in stellar mass with these analyses and falls off below 0.5\mearth, although there is at least one KOI hosted by a mid-M dwarf just above the radius valley (Fig. \ref{fig:rad_mk}).

\begin{figure}
\centering
\includegraphics[width=\columnwidth]{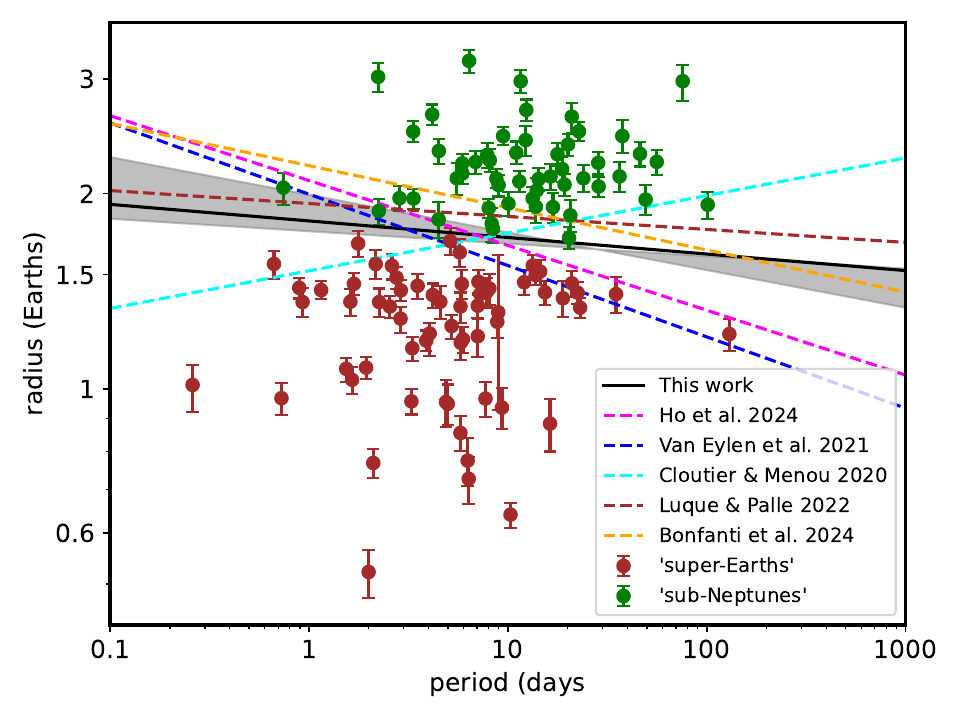}
\caption{Linear boundary (black line) between 'super-Earths' and `sub-Neptunes' in log period-log radius space derived by support vector machine regression, using the full sample.  Planets are color-coded according to the self-consistent (iterated) SVM classification and the grey region represents the 68\% confidence range based on $10^4$ Monte Carlo bootstrap representations of the data.  Regressions of the valley derived for M dwarf host stars by other investigations are represented by the colored dashed lines.  The \citet{Ho2024} slope is the mass-and period-dependent equation $\log R_p = -0.10 \log P + 0.17 \log M_* + 0.36$ evaluated at the median stellar mass of our sample of 0.59\msun.}
\label{fig:valley}
\end{figure}

The location of the radius valley and its dependence on orbital period (a proxy for irradiance) is used to test models of escape of planetary atmospheres/envelopes (see Sec. \ref{sec:discussion}).   Following past work \citep[e.g.,][ and references therein]{Ho2023}, we performed a regression on the full sample with support vector machines (SVM)  as implemented in the {\tt Python} {\tt sci-kit learn} package, with a choice of regularization parameter of $C = 10$, following \citet{VanEylen2019}.  Super-Earth and sub-Neptune classification was initially assigned initially based on radii smaller or larger than 1.68\rearth{} (the value predicted for this sample from the relation in \citet{Ho2024}), then classification was iteratively redone (10 times) using the SVM.  Confidence intervals were calculated with $10^4$ Monte Carlo simulations where planet radii were randomized according to their uncertainties.  We plot the final assignments and converged regression ($\log R_p = -0.03^{+0.01}_{-0.03} \log P + 0.26^{+0.04}_{-0.01}$) as the black line in Fig. \ref{fig:valley}, along with previously published regressions.  

Our regression is only weakly period-independent and contrasts with the negative slopes for M dwarf host stars derived by \citet[][blue dashed line in Fig. \ref{fig:valley}]{VanEylen2021}, from combined solar- and late-type \kepler{} host stars derived by \citet{Ho2024} and evaluated at the 0.59\msun{} median sample mass of our sample (dashed magenta line), and for a sample of \tess{}-detected planets around M dwarfs  with mass measurements \citep[][dashed orange line]{Bonfanti2024}.  It is also inconsistent with the positive slope found by \citet{Cloutier2020} for planets detected around late-type hosts by \kepler/\ktwo{} (dashed cyan line).  However, the slope is consistent with that derived from a small sample of M dwarf planets with RV-determined masses by \citet{Luque2022}.  The cause of the disagreements between these estimates is not known, but could arise from sensitivity of the slope to objects with the shortest or longest orbital periods combined with refinement of radii by exclusion of binaries and improved stellar properties.

\subsection{Corrected Radius Distribution}
\label{sec:corrected-radius}

\begin{figure}
	\includegraphics[width=\columnwidth]{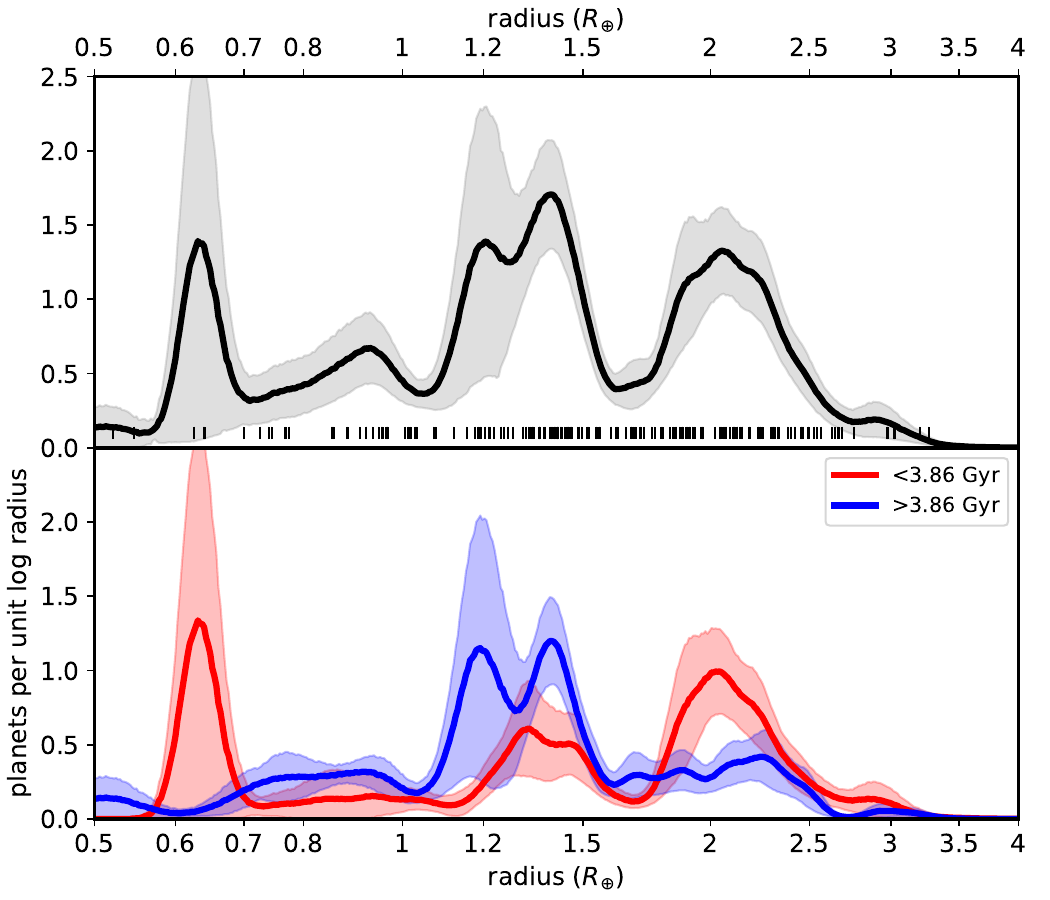}
    \caption{Empirical, non-parametric radius distribution of planets hosted by late K- and early M-type dwarfs observed by \kepler{} and assigned ages based on rotation, corrected for detection efficiency.  The top panel shows the distribution of the entire sample, with radii of individual objects denoted by tick marks.   The bottom panel show the distributions for objects with assigned ages partitioned into equal younger ($<3.86$ Gyr) and older halves.  Shaded regions are 68\% confidence interval determined from Monte Carlo sampling with replacement.}  
    \label{fig:radius}
\end{figure}

The detection efficiency-corrected radius distribution constructed for the entire small-planet sample is shown in the top panel of Fig. \ref{fig:radius}.  The grey region spans the 68\% confidence interval, and the individual planet radii are marked by the ticks at the bottom.    The total occurrence is the integral over the distribution and is 1.84 planets per star.   This presentation also shows the super-Earths and sub-Neptunes sa peaks separated by an intervening gap or ``valley" around 1.6-1.7\rearth.  We computed a valley depth, i.e. the ratio of the mean of the adjacent maximum to the minima, of 3.85.  This is similar to that found for solar-type stars by \citet{Ho2023}, and inconsistent with the trend of shallower valley with lower stellar mass found by \citet{Ho2024}.        

While Fig. \ref{fig:radius} appears to resolve the super-Earth population into two overlapping peaks, this structure is not significant compared to the uncertainties in the reconstruction (represented by the shaded area).  To verify this, for each of the 1000 bootstrap reconstructions we assessed whether the radius interval 1-1.7\rearth{} contained one or two maxima and, in the latter case, calculated the ratio of the intervening minimum to the lowest of the two maxima.  75\% of the reconstructions contain two maxima (at $\approx$1.2 and 1.4 \rearth), but among these, the ratio of the minimum to lowest maximum is continuously distributed between about 0.6 to unity, with most values near unity.  This indicates that the two observed peaks are a random occurrence from a continuous distribution of possible outcomes. 

There are other, more ambiguous features in the radius distribution, including a ``shoulder" at 3\rearth, representing four planets of similar size (Fig. \ref{fig:period-radius-kde}).   The radius distribution also hints at the existence of a population of Earth-size objects distinct from super-Earths (second peak from left in Fig. \ref{fig:radius}).  This is not an artefact of detection bias, since the efficiency only decreases smoothly with radius in this regime (Fig. \ref{fig:period-radius-kde}).  The prominent peak at 0.64\rearth{} in Fig. \ref{fig:radius} is produced by KOI--314.03 (Kepler--138b), which has a low detection probability due to its comparatively wide (10.3 day) orbit.  It is a member of a three-planet system and was discovered only after analysis with optimized, automated routines following the discovery of the other, larger members \citep{Batalha2013}.  Its presence in the sample could be just a statistical fluke, or it could represent a large population of ``sub-Earths" \citep{Sinukoff2017}.  
 
\subsection{Radius Evolution}
\label{sec:rad_evo}

We divided the planets with assigned ages into halves that are younger and older than the median gyrochronologic age of 3.86 Gyr (mean ages of 2.6 and 5.4 Gyr, respectively).   These are plotted red and blue curves in the bottom panel of Fig. \ref{fig:radius}.  The sub-Neptune peak is much less prominent in the older sample while the super-Earth is larger, with the ratio between the two (integrated over 1.68--4\rearth{} and 0.7--1.68\rearth, respectively) declining from 2.38 to 0.58 from the younger to older moieties.  This is consistent with the evolution of planets by the loss and/or contraction of low molecular-weight envelopes over Gyr.  Indeed, the integrated probabilities over mini-Neptunes (1.7--4\rearth) fall from 0.59 to 0.31 and is roughly equivalent to the increase in the Earth- and super-Earth (0.7--1.7\rearth) from 0.25 to 0.53.  There is no evidence for a decrease in the mean radius of the sub-Neptunes with age nor a deepening of the radius valley as has been seen in other analyses of solar-type stars \citep{Petigura2022,ChenD-C2022} but the valley is indistinct in the older sample due to the small number of sub-Neptunes.  A two-sided Kolmogorov-Smirnov test provide weak statistical support for the young and old samples to be drawn from different populations ($p = 0.08$).

We investigated four systematic effects that could potentially mimic radius evolution: (1) an artefact due to a sensitivity of the radius distribution to the choice of divisor between younger and older sub-samples; (2) a variation in stellar properties with age that influences detection efficiency; (3) a difference in host star properties brought about by selecting for stars with detectable rotation period that in turn affects the detected planets; and (4) a variation in stellar properties (i.e. \teff{} or mass) with age combined with a correlation between the planet population and stellar properties.  We find no evidence that any of these is responsible for the observed difference in radius distribution between the younger and old sub-samples, and details are provided in Appendix \ref{sec:systematics}.  

Figure \ref{fig:period-radius} compares the period-radius distribution of the younger (red points) and older (blue points) sub-samples and hints at non-uniformity in this evolution.  While there are both younger and older sub-Neptunes ($>1.7$\rearth, above the horizontal line) at periods $<25$ days, nearly all those at longer periods ($>$25 days) are younger.  This is discussed further in Sec. \ref{sec:cool}.  

\begin{figure}
    \centering
    \includegraphics[width=\columnwidth]{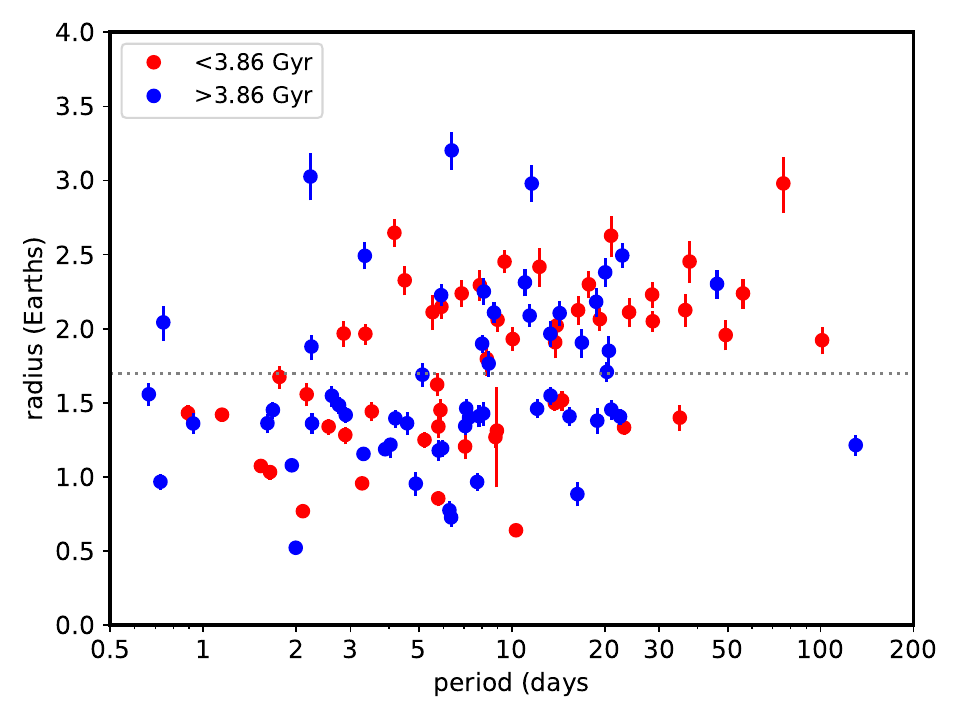}
    \caption{Radius vs. orbital period of planets in systems with ages younger (red points) or older (blue points) than the median age of the sample (3.86 Gyr).  The grey contours map the \kepler\ detection efficiency (including the geometric factor) for the DR25 catalog with late K and early M dwarf host stars of all ages, i.e. the fraction of planets detected if every star in the sample hosted a planet of a given radius and orbital period.}
    \label{fig:period-radius}
\end{figure}

\section{Interpretation and Discussion}
\label{sec:discussion}

\subsection{The Radius Valley}
\label{sec:valley}

Models that explain the dichotomy between super-Earths and sub-Neptunes predict a dependence of the location of the radius valley on orbital period or stellar irradiation similar that can be compared to observations \citep[e.g., see Table 6 in ][]{Ho2023}.  Both photoevaporation and core-powered mass loss models predict negative slopes with period, in the former due to stellar X-ray and/or EUV radiation as the driver \citep[e.g.,][]{Owen2012}, and in the latter a consequence of the role of stellar bolometric irradiation \citep{Gupta2021}.   For example, \citet{Gupta2019} derived an index of -0.11 for core-powered loss, independent of stellar mass, while \citet{Owen2017} derive -0.16 for photoevaporative loss.  Such models are consistent with the slopes reported by \citet{VanEylen2021}, \citet{Ho2024}, and \citet{Bonfanti2024} (Fig. \ref{fig:valley}).  In contrast, a model of planet formation in a gas-depleted inner disk predicts a \emph{positive} slope of +0.11 \citep{Lopez2018}, consistent with the findings of \citet{Cloutier2020}.  None of these are consistent with the near-zero slopes found by \citet{Luque2022} and our analysis ($-0.03^{+0.01}_{-0.03}$; Fig. \ref{fig:valley}).  Weak dependence on period as well as on stellar luminosity  (Fig. \ref{fig:rad_mk}), suggests a mechanism largely independent of the host star, e.g. differences in water abundance \citep[see ][for a discussion of some possible scenarios]{Burn2024}.  

The depth of the radius valley relative to the adjacent super-Earth and sub-Neptune peaks has also been compared to model predictions.  In particular, \citet{Ho2024} found that the valley became shallower (filled in) at lower host star masses.  A trend could appear due to error in the radius ratio $R_p/R_*$ from fitting the transit lightcurve, which, all else being equal, is expected to be higher for shorter duration transits around smaller M dwarf host stars \citep{Petigura2020}.  Astrophysical interpretations of a shallower radius valley include: (a) scatter in the XUV-emission luminosity of young stars at a given mass and age; (b)  heterogeneity in the composition of planetary cores; and (c) collisions \citep{Ho2024}.  Our failure to reproduce the trend motivates a deeper dive into the systematics that could affect reconstructions of the planet radius distribution.

\subsection{Evolution of Sub-Neptunes into Super-Earths}
\label{sec:evolution}

The decline of the sub-Neptune population and commensurate rise in super-Earths between our younger and older sub-samples (Fig. \ref{fig:radius}) indicates that some sub-Neptunes have evolved into super-Earths over a timescale of $\sim$3 Gyr.  On orbits with $P > 20$ day this radius evolution has presumably caused some planets to leave the sample due to the low detection efficiency of super-Earths on wider orbits  (grey contours in Fig. \ref{fig:period-radius-kde}).  The observed change is analogous to that seen around solar-type \kepler\ host stars \citep{Berger2020,Sandoval2021,David2021,ChenD-C2022}, which has been explained by a scenario where sub-Neptunes are super-Earths with thick hydrogen-dominated envelopes which can escape by XUV-driven photoevaporation or core-powered mass-loss from rocky/icy cores \citep{Owen2019,Gupta2021,Owen2024}.  

High-energy emission from solar-type host stars is most intense during the first few hundred Myr \citep{Johnstone2021}, so a Gyr timescale for planet evolution has been previously interpreted as supporting core-powered  \citep{Berger2020}.  However, more quantitative estimates of photoevaporative timescales are longer \citep{King2021} and both mechanisms may operate at different times in different regimes \citep{Owen2024}. Moreover, our sample cannot be used to investigate shorter timescales because it does not include systems with ages much younger than 1 Gyr.  This is because low mass stars take longer to converge to an initial condition-independent rotation sequence, where gyrochronology can be applied.

In a scenario where sub-Neptunes are rocky cores with thick envelopes of high molecular weight volatiles, specifically \water{} \citep{Turbet2020}, radius evolution could be at least partly explained by contraction due to cooling \citep{Howe2015,Vazan2018b}, and/or chemical reaction between the envelope and the rocky core \citep{Kite2020,Kim2023}.  These processes have timescales different from that of atmospheric escape mechanisms.

\subsection{The Cool Sub-Neptune Collapse}
\label{sec:cool}

The near-complete absence of sub-Neptunes in the ``desert" very close to the host stars is consistent with the canonical picture of envelope loss driven by stellar irradiation.  However, the disappearance of sub-Neptunes elsewhere appears to be non-uniform (Fig. \ref{fig:period-radius}).  There is a striking paucity of older sub-Neptunes on the widest orbits ($>$20 days), where irradiation is lowest and escape should be slower than at intermediate orbits (2--20 days) where older sub-Neptunes are found (Fig. \ref{fig:period-radius}).    

\begin{figure}
    \centering
    \includegraphics[width=\columnwidth]{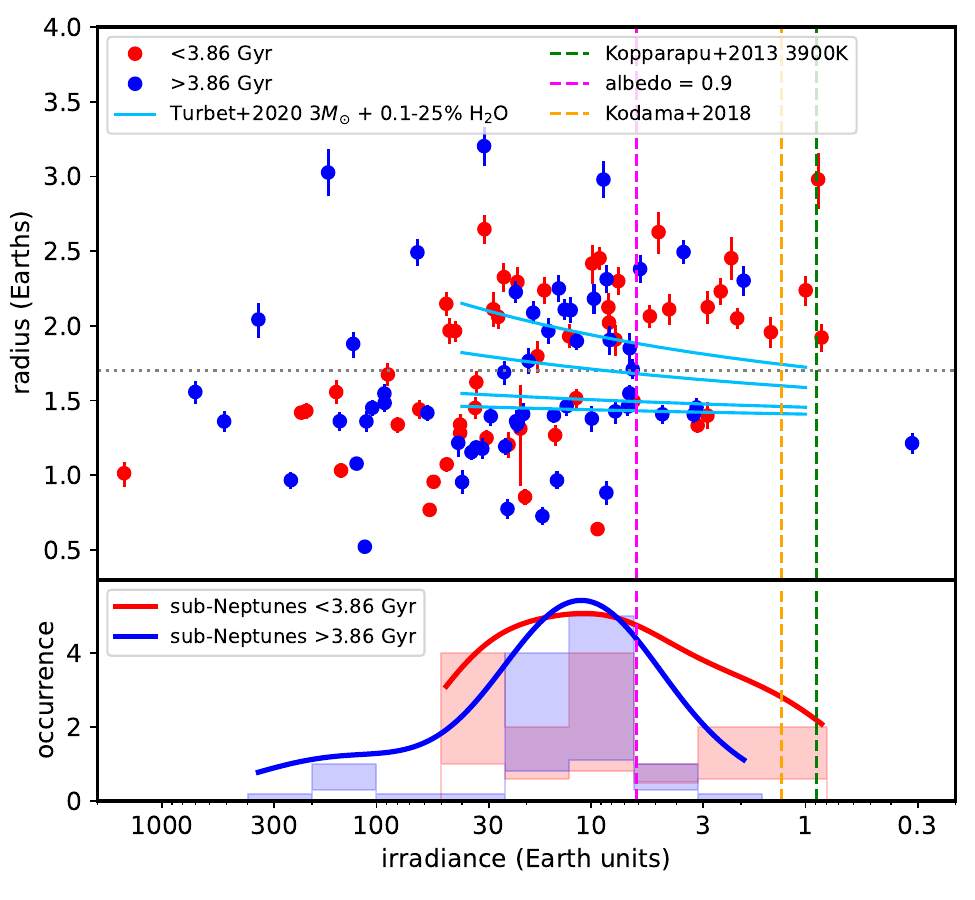}
    \caption{Top: Radius vs. total stellar irradiance or ``instellation" (in units of the current terrestrial value) for planets in systems with ages younger (red) or older (blue) than the median age of the sample (3.86 Gyr).  Horizontal error bars are smaller than the points.  Vertical lines mark different irradiance thresholds for a runaway water vapor greenhouse atmosphere above which a planet's atmosphere is predicted to be inflated.  Cyan lines mark predicted radii of planets with a 3\mearth\ rocky core and (bottom to top) 0.1, 1, 10, and 25\% H$_2$O mass fractions \citep{Turbet2020}.  Bottom: Relative distribution of younger (red) and older (blue) sub-Neptunes (1.7-4\rearth) with irradiance.  Shaded bars present histograms spanning the 68\% probability range based on Poisson statistics.  The solid curves were generated by Gaussian kernel density estimators with Scott's bandwidth \citep{Scott1992}.  }
    \label{fig:radius-flux}
\end{figure}

The top panel of Figure \ref{fig:radius-flux} shows planet radius versus stellar irradiance, comparing the younger and older sub-samples.  Irradiance is based on bolometric luminosities calculated using interpolations of the corrections to $K$-band luminosities tabulated in \citet{Pecaut2013}, and assuming near-circular orbits.  Constructed irradiance distributions for younger and older sub-Neptunes ($R_p > 1.7$\rearth, dotted grey line) are shown in the bottom panel.  Both Gaussian kernel density estimators (solid curves) and histograms with Poisson statistics errors (shaded bars) suggest a deficit of older sub-Neptunes at irradiances below several times the present terrestrial value (\searth).  The fraction of sub-Neptunes in the older-sub-sample falls from 53\% at irradiances $>$5\searth\ to 17\% at lower irradiance (see red and blue histograms).   This cannot be due to a differences in detection efficiency since planets on wider orbits should be more readily detected around older, photometrically quieter stars.  Nor is there evidence for a combination of correlations between age with host star mass (or luminosity), and stellar mass with planet period distribution (Figs. \ref{fig:age_mk} and \ref{fig:period-mk} in Appendix \ref{sec:systematics}) that could mimic this effect.  There are no sub-Neptunes in the sample below an irradiance of 1\searth, although this could be due in part to low detection efficiency at long orbital periods.  The two planets that are marginally that are least irradiated are KOIs 812.03 (Kepler--235e) and 952.03 (Kepler--32d), with ages of $7.8 \pm 1.43$ and $5.60 \pm 0.96$ Gyr, respectively \citep{Gaidos2023b}. 

The persistence of ``warm" sub-Neptunes and disappearance of cool sub-Neptunes at older ages could be due to an inflation mechanism that depends on stellar irradiation or planet equilibrium temperature and hence proximity to the host star.  \citet{Turbet2020} showed that primarily rocky planets with water-dominated atmospheres could have sub-Neptune-like radii at irradiances exceeding a ``runaway" threshold where \water{} cannot condense into a liquid ocean.  

\citet{Kopparapu2013} and \citet{Turbet2019} found the irradiance threshold for a runaway greenhouse on Earth-like planets to be $\approx$1.05\searth.  Around M dwarfs stars the threshold is slightly lower because of reduced Rayleigh scattering and increased absorption by atmospheric H$_2$O at the longer wavelengths where cooler stars emit more radiation; the value is predicted to be 0.9\searth\ for \teff=3900K or M0 spectral type (dashed vertical green line in Fig. \ref{fig:radius-flux}.     Three-dimensional general circulation models find a somewhat higher threshold that depends on planet rotation rate \citep[][; vertical orange dashed line]{Leconte2013,Kodama2018}.  The higher Bond albedo of a cloudy steam atmosphere (that of Venus is around 0.76 \citep{Moroz1985} but measurements of the phase function suggest a value as high as 0.9 \citep{Mallama2006}\footnote{The albedo depends would depend strongly on cloud droplet size and the high reflectivity of Venus is largely due to sulfuric acid aerosols and a similar scenario might not apply to steam atmospheres.}), might move the runaway threshold to $\sim5$\searth.  This is the approximate boundary where sub-Neptunes become persistent (vertical dashed magenta line).  Finally, atmospheres containing significant H/He will have a lower threshold \citep{Innes2023}.      

While a decline in sub-Neptune occurrence across the water condensation boundary could, in principle, be explained by the collapse of steam atmospheres, it is unclear if \water{} alone can explain the observed distributions with radius and age.   In Fig. \ref{fig:radius-flux} we plot predictions of \citet{Turbet2020} for \water{} mass fractions between 0.1 and 25\% (cyan lines), assuming a 3 \mearth{} rocky core mass chosen to reproduce super-Earth radii.  This comparison suggests that \water{} mass fractions exceeding 25\% are necessary to explain some sub-Neptunes.   Models of \water-rich planets produce radii up to 3\rearth{} \citep{Burn2024}, but \citet{Vivien2022} found that pure water/steam atmospheres on sub-Earths become unstable if the atmospheric scale height exceeds 0.1 times the planet radius.  Moreover, some internal governing mechanism \citep[e.g., reaction with a magma ocean,][]{Schaefer2016} is needed to explain the radius evolution over Gyr since the change in luminosity of these very low-mass host stars is small and the response of an atmosphere nearly instantaneous. 

Alternatively, the inflated radii of sub-Neptunes and the differential evolution of cool vs. warm objects could be explained by atmospheres of mixed H/He and \water{}.   H/He lowers the mean molecular weight and decreases the irradiance threshold for a runaway water-vapor greenhouse \citep{Innes2023}.  Slow escape of H/He would eventually collapse sub-Neptunes to smaller objects.  At an elevated irradiances above the \water{} condensation threshold, \water{} could be persist in the upper atmosphere where its radiative cooling the lowers temperature and inhibits H/He escape \citep{Yoshida2022}.  In a cooler sub-Neptune, water condenses into clouds deeper in the atmosphere and cannot play that role.   The upper atmosphere is therefore hotter, H/He escape faster, and the sub-Neptunes eventually collapses to a rocky core with a geometrically thin atmosphere.

\begin{figure}
    \centering
    \includegraphics[width=\columnwidth]{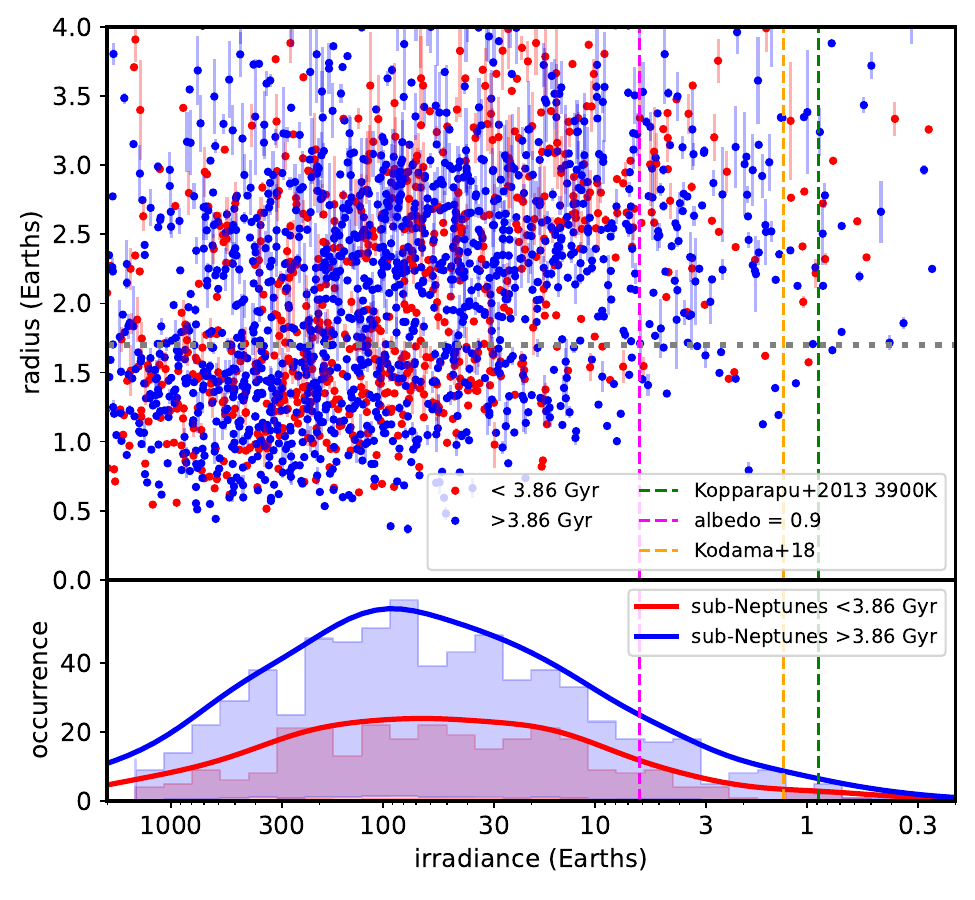}
    \caption{Top: Distributions of planet radius vs. stellar irradiation for systems around solar-type stars with ages that are younger (red points) or older (blue points) than 3.86 Gyr, from \citet{Berger2018b}.  Bottom: Relative distribution of younger (red) and older (blue) sub-Neptunes (1.7-4\rearth, above the horizontal grey dotted line in the top panel) with irradiance.  Shaded bars present histograms spanning the 68\% probability range based on Poisson statistics.  The solid curves were generated by Gaussian kernel density estimators with Scott's bandwidth.  Compare to Fig. \ref{fig:radius-flux} for late K and early M-type stars.  Here, systematic errors in irradiance are likely larger than counting statistics.}
    \label{fig:berger}
\end{figure}

Finally, we searched for irradiance-dependent evolution of sub-Neptunes among \kepler\ transiting planets around solar-type host stars with isochrone ages determined by \citet{Berger2018b}, restricting the sample to those with ages more precise than 30\%.  Although the median age of the sample is 5 Gyr, we use the same division of 3.86 Gyr between young and old populations for comparison. No such effect is obvious in the irradiance distributions (Fig. \ref{fig:berger}).  The ratio of young to old planets is statistically indistinguishable ($\approx0.44$) for planets experiencing $S > 5$\searth\ and $S < 5$\searth.  Three explanations for the difference between the two samples are (1) the late K/early M dwarf sample is affected by small-number statistics and is a statistical fluke; (2) the ages of the \citet{Berger2018b} sub-sample with relatively cool planets are more uncertain because the host stars are likely to be late G and early K dwarfs which evolve very slowly in a color-magnitude diagram; and (3) there is a fundamental difference in the evolution of the two planet populations.  At a given bolometric irradiance, planets around M dwarfs will experience far more X-ray and UV irradiation than their solar-type counterparts at \citep{McDonald2019} and it is this radiation that drives photoevaporation of H/He atmospheres \citep{Owen2019}.  Thus temperate sub-Neptunes around solar-type stars will have retained H/He atmospheres while their counterparts around M dwarfs have lost most or all their H/He and have radii controlled by the behavior of water-dominated atmospheres.  

\begin{figure}
    \centering
    \includegraphics[width=\columnwidth]{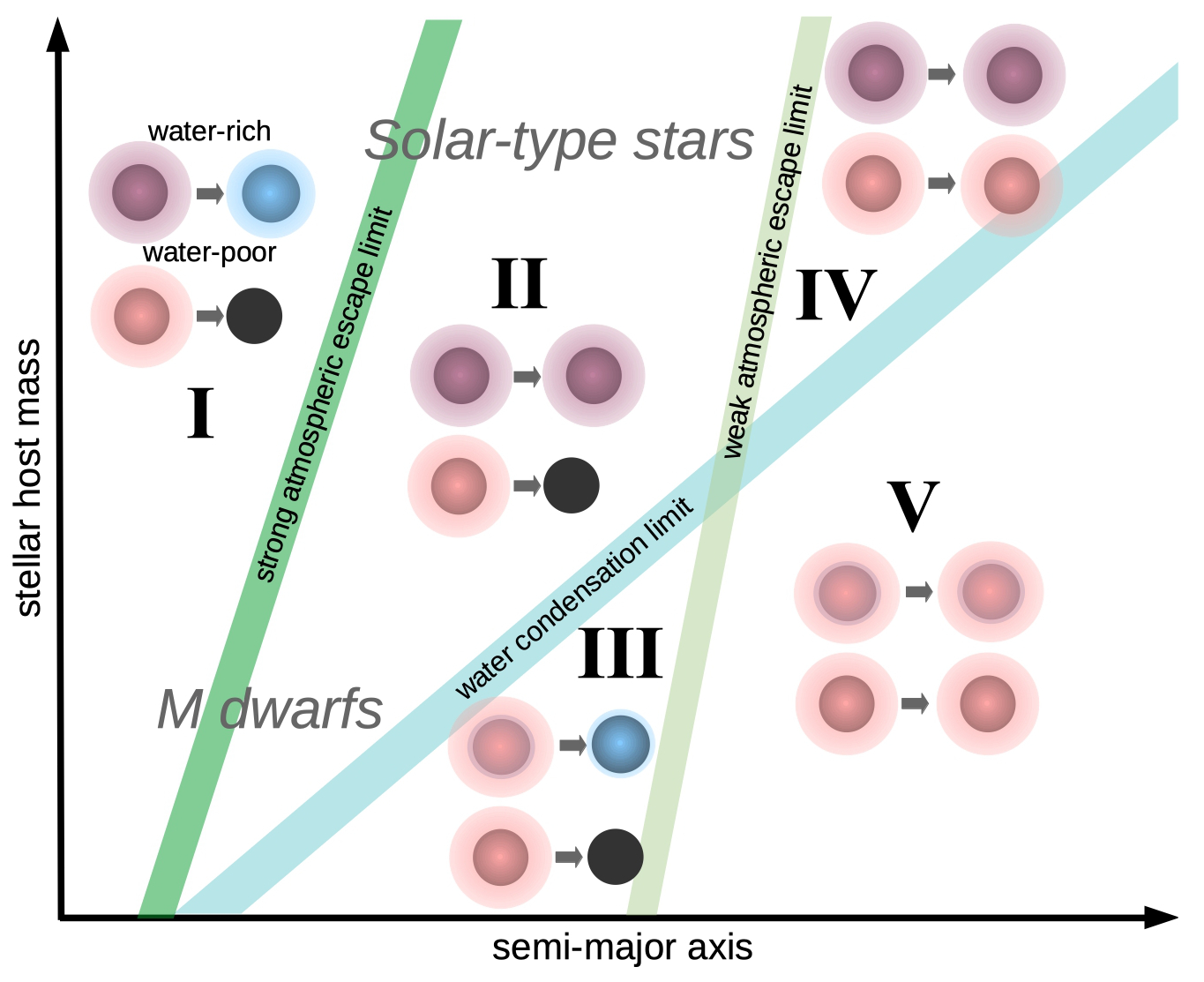}
    \caption{A pictorial schematic of zones in the orbit- and stellar-mass-dependent evolution of sub-Neptunes.  Boundaries between the zones are merely illustrative and not to scale.  Within each zone, the upper and lower pathways represent water-rich and water-poor planets, respectively.  Red represents hydrogen/helium, blue represents \water, and black represents rocky cores.  Interior to the water condensation limit (blue boundary), \water{} is present in the upper atmospheres of water-rich planets.  Exterior to the strong atmospheric escape limit, atmospheric water inhibits escape of H/He.  Exterior to the weak atmospheric escape limit, escape of H/He is negligible.  The difference in slopes represent the greater sensitivity of bolometric luminosity to stellar mass compared to XUV luminosity \citep[e.g.,][]{McDonald2019}.}
    \label{fig:cartoon}
\end{figure}

These different pathways are illustrated in Fig. \ref{fig:cartoon}, where the \kepler{} M dwarf sample spans Zones I, II, and III (but not Zone V due to small sample size), while the \citet{Berger2018b} sample spans Zones I, II, and IV.  In Zone I, rapid stellar-driven escape of H/He is not suppressed by atmospheric \water{} and leads to planets with extended steam-rich atmospheres or rocky super-Earths (Fig. \ref{fig:cartoon}).  In Zone II, water-rich (but not water-poor) sub-Neptunes retain thick H/He envelopes.  In Zone III (only around very low-mass stars), H/He envelopes are lost, leaving ``ocean worlds" and rocky super-Earths, and in contrast to Zones 4 and 5, where these envelopes are retained.  Measurements of planet mass by RVs and observations of atmospheres by \jwst\ for examples around much closer, brighter host stars could distinguish between these scenarios. 

\subsection{Caveats and Limitations of this Analysis}
\label{sec:caveats}

Biases and selection effects that could possibly mimic the observed radius evolution are summarized in Sec. \ref{sec:rad_evo} and evaluated in Appendix \ref{sec:systematics}.  There are other effects that could quantitatively impact our results.  We restricted our sample to planet host stars with detected rotational signals in \kepler\ lightcurves and these are a biased subset of the overall population. Younger stars will be rapidly rotating, magnetically active, and highly spotted and their rotational periods can be readily measured.  However, due that rotational variability as well as intense flaring can inhibit the detection of transiting planets \citep{Gaidos2017a}.  On the other hand, the oldest stars will be slowly rotating, less spotted, and exhibit less rotational variability, and the rotation periods can be similar to the \kepler\ quarterly roll ($90$ days) and thus more difficult to retrieve \citep{Santos2019}.  This is particularly true for the coolest stars in the sample (mid M-dwarfs) because these stars have the longest rotation periods for a given age.  These selection effects can manifest themselves in the overall distribution of M dwarf planet host ages \citep[e.g.,][]{Gaidos2023b}.  However, we see no trend of planet properties (i.e., radius) with host star properties (i.e., luminosity, Fig. \ref{fig:rad_mk}).  

The overall occurrence rate estimated from KOIs with rotation periods could be biased upwards to the extent that stellar rotation and planetary orbits are aligned \citep[low stellar obliquity][]{Albrecht2022}.  Stars viewed at a high inclination (equator on) will have the strongest rotational signal and their planets will be more likely to transit compared to the overall population.  We quantified this effect by determining the fractional numbers of KOIs around K/M dwarfs observed by \kepler{} with rotation periods in \citet{Santos2019} vs. K/M dwarfs lacking rotation periods.  We applied the same cuts ($M_K > 4.5$, RUWE $<1.4$, and not otherwise identified as binaries) to both samples, but we did not apply a \teff{} cut since temperatures are only available for the \citet{Santos2019} sample.  The KOI fractions are  $2.70 \pm 0.24$\% and $2.50 \pm 0.39$\%, respectively, with uncertainties from counting statistics.  These are thus indistinguishable and we conclude the effect is negligible among these types of stars, perhaps because of the high (\%73) success rate of detecting rotation among these stars.

Transit parameters and stellar parameters were not computed self-consistently: the former adopted DR25 values and the latter were based on revised luminosities and new empirical relations.  Stellar density is covariant with transit impact parameter which, along with \teff\ (which controls limb darkening coefficients) is covariant with the planet-to-star radius ratio $R_p/R_*$.  Lastly, our filtering of binaries from the \kepler\ late K- and early M-dwarf target sample was less thorough because AO imaging is not available for the overall \kepler\ late K/early M dwarf population as it is for the KOIs.

\section{Summary and Future Directions}
\label{sec:summary}

We refined the radii of 117 transiting planets detected by \kepler{} around 74 cool (3200--4200K) single dwarf stars and assigned ages to 113 based on their rotation and a revised gyrochronology.  Based on these values, we conclude that: (1) analogous to the distribution found for solar-type host stars, the detection efficiency-corrected radius distribution of M dwarf planets contains two peaks representing super-Earths and sub-Neptunes, with the sub-Neptunes largely absent in a ``desert" at periods $<2$ days; (2) in contrast to some previous investigations of both solar-type and cooler dwarfs, the planet radius valley separating these two populations is only weakly dependent on orbital period (d$\log R_p$/d$\log P = -0.03^{+0.01}_{-0.03}$); (3) there is a decline in sub-Neptunes and rise in super-Earths over the $\sim$3 Gyr separating the younger and older halves of the sample; and (4) the decline in sub-Neptunes is mostly at lower stellar irradiances where water can condense as clouds deep in the atmosphere.  The last effect is not seen among \kepler{} planets around solar-type stars where the water condensation limit is much more distant.  Taken together these suggest a governing role for intrinsic compositional differences in the radius dichotomy of small planets e.g., planets having water-rich and water-poor envelopes lying above and below the valley, respectively.  The relative decline in cooler sub-Neptunes relative to their warmer counterparts with time could be due to cooling of the upper atmosphere by water and inhibition of escape of H/He.  

Our findings, while intriguing, are limited in their statistical robustness by sample size rather than the precision with which stellar and planet parameters can be determined.  The small number of late K and M dwarf KOIs is the result of the selection criteria for targets of the \kepler\ prime mission and the telescope's single fixed pointing.  Both the \ktwo\ and \tess\ mission have identified many more transiting planet systems around such stars, some of which have ages assigned by gyrochronology \citep{Gaidos2023b}.  However, determining the rotation periods of late-type field stars from \ktwo\ and especially \tess\ photometry alone is challenging \citep{Reinhold2020,Claytor2022}.  In the \citet{Gaidos2023b} catalog of late K- and early M-type stars with established rotation periods and ages there are only 18 \tess{} Objects of Interest (TOIs).  This situation will improve with the discovery of more TOIs hosted by very cool stars, plus long-period planets near the water-condensation limit detected with only one or two transits by \tess{} and confirmed by other observatories \citep{Gill2020,Canas2020,Yao2021,Magliano2024}.  However, obtaining rotation periods of middle-aged cool host stars for gyrochronology will in most cases  require sources of photometry other than \tess{} or analysis of activity indicators in the spectra obtained for radial velocity monitoring.  The \emph{PLATO} mission and its long-duration (two-year) fields could yield additional systems with gyrochronologic ages, depending on target selection \citep{Nascimbeni2022}.  

The water condensation limit should be marked by a transition in the abundance of H$_2$O in the upper atmospheres of cool sub-Neptunes, a boundary that is beginning to be explored by infrared transmission spectroscopy by \jwst{}, e.g. TOI-270d \citep{Holmberg2024}.   There are also additional investigations that could be performed with larger samples, for example, the expected dynamical evolution of multi-planet systems due to mutual perturbations \citep{Teixeira2023}.\footnote{The orbital period distributions of the young and old sub-samples presented here do not appear significantly different (K-S test $p = 0.15$).}

\begin{table}
\label{tab:kois}
\caption{\kepler{} M Dwarf Planet Radii and Ages$^{1}$}
\include{table1.tex}
\begin{flushleft}
$^{1}$The full table is available as a machine-readable table (http://zenodo.org).\\
\end{flushleft}
\end{table}
%\end{landscape}

\section*{Acknowledgements}

EG and AA acknowledge support from the NASA Exoplanet Research Program (Award 80NSSC20K0957). ALK acknowledges support from the NASA Exoplanet Research Program (Award 80NSSC22K0781). Some of the data presented herein were obtained at the W. M. Keck Observatory, which is operated as a scientific partnership among the California Institute of Technology, the University of California and the National Aeronautics and Space Administration. The Observatory was made possible by the generous financial support of the W. M. Keck Foundation.   This research has made use of the NASA Exoplanet Archive, which is operated by the California Institute of Technology, under contract with the National Aeronautics and Space Administration under the Exoplanet Exploration Program.  This work has made use of data from the European Space Agency (ESA) mission
{\it Gaia} (\url{https://www.cosmos.esa.int/gaia}), processed by the {\it Gaia}
Data Processing and Analysis Consortium (DPAC,
\url{https://www.cosmos.esa.int/web/gaia/dpac/consortium}). Funding for the DPAC
has been provided by national institutions, in particular, the institutions
participating in the {\it Gaia} Multilateral Agreement.

 \section*{Data Availability} 
 
The KIC and DR25 KOI catalogs are available from the, \gaia\ DR3 astrometry and photometry are available from the CDS, and AO imaging is publicly available from the Keck Observatory Archive.\\

%%%%%%%%%%%%%%%%%%%%%%%%%%%%%%%%%%%%%%%%%%%%%%%%%%



\appendix
\section{Biases and Selection Effects}
\label{sec:systematics}

We evaluated potential biases and selection effects that could cause the reconstruction of the planet radius distribution to mimic the observed evolution. 

\begin{figure}
	\includegraphics[width=\columnwidth]{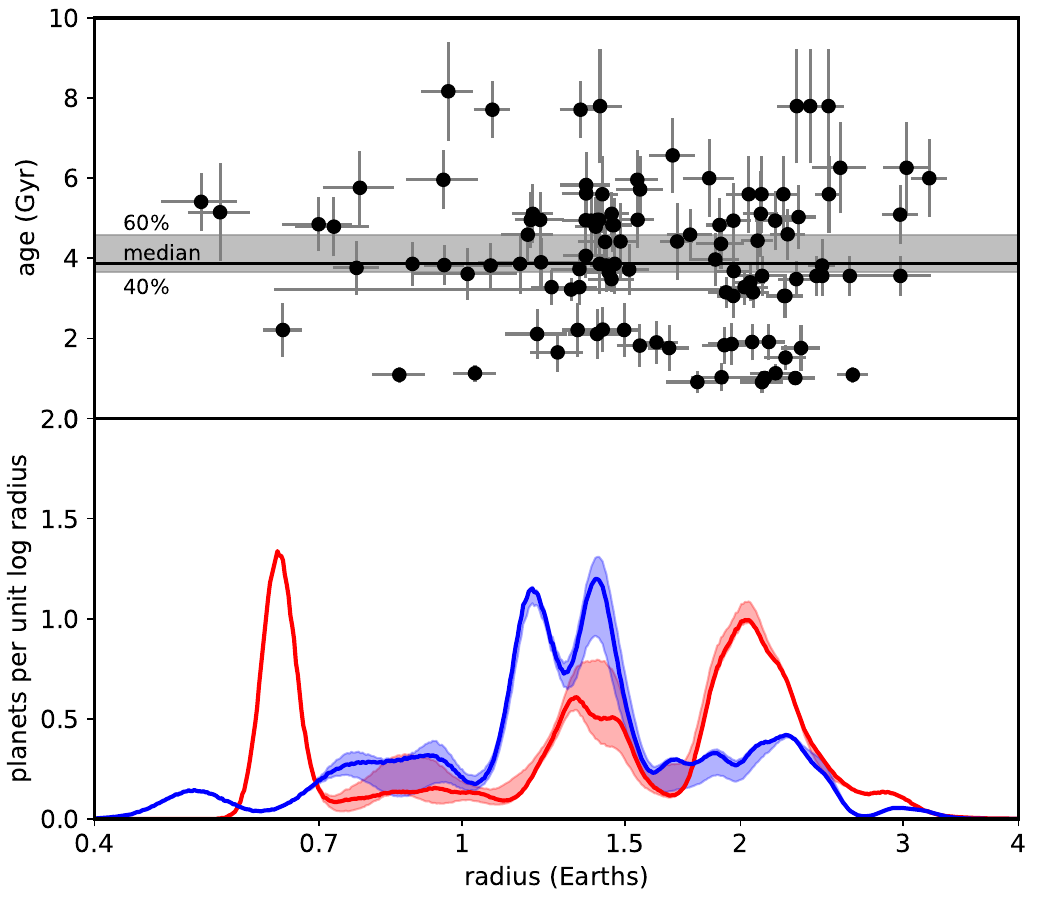}
    \caption{Sensitivity of the change in radius distribution to the divisor between young and old samples.  Top: Planet radius vs. gychronologic system age.  Three possible age cuts between young and old stars are shown; the median age (3.86 Gyr) is represented by the solid line, and 40 and 60 percentile ages (3.65 and 4.59 Gyr) bound the grey region.   Bottom: Radius distributions of young (red) and old (blue) planets using the three divisor ages, with those using the median represented by the solid line, and those constructed using the 40 and 60 percentile values spanning the shaded regions.}  
    \label{fig:age-radius}
\end{figure}

\emph{Choice of age divisor:} We investigated the sensitivity of the difference in radius distribution between young and older systems to the choice of age to divide the sub-samples.  The choice of median (3.86 Gyr, horizontal line in the top panel of Fig. \ref{fig:age-radius}) is a statistical convenience and gyrochronologic ages have significant uncertainties which can ``blur" age dependence.  To assess this, we generated alternative radius distributions, dividing the sample a the 40 and 60 percentile values of the age distribution (3.65 and 4.59 Gyr).  These bound the grey shaded region in the top panel of Fig. \ref{fig:age-radius}.  In the bottom panel, the radius distributions for the younger and older sub-samples are shown as the red and blue curves, respectively, with the solid curves those constructed using the median age and the shaded regions bounded by the results from the 40 and 60 percentile values.  All younger distributions exhibit an elevated number of sub-Neptunes to super-Earths whereas for older distributions the reverse is true.  Thus the evidence for radius evolution is robust to the exact choice of sample divisor.    

\begin{figure}
    \centering
    \includegraphics[width=\columnwidth]{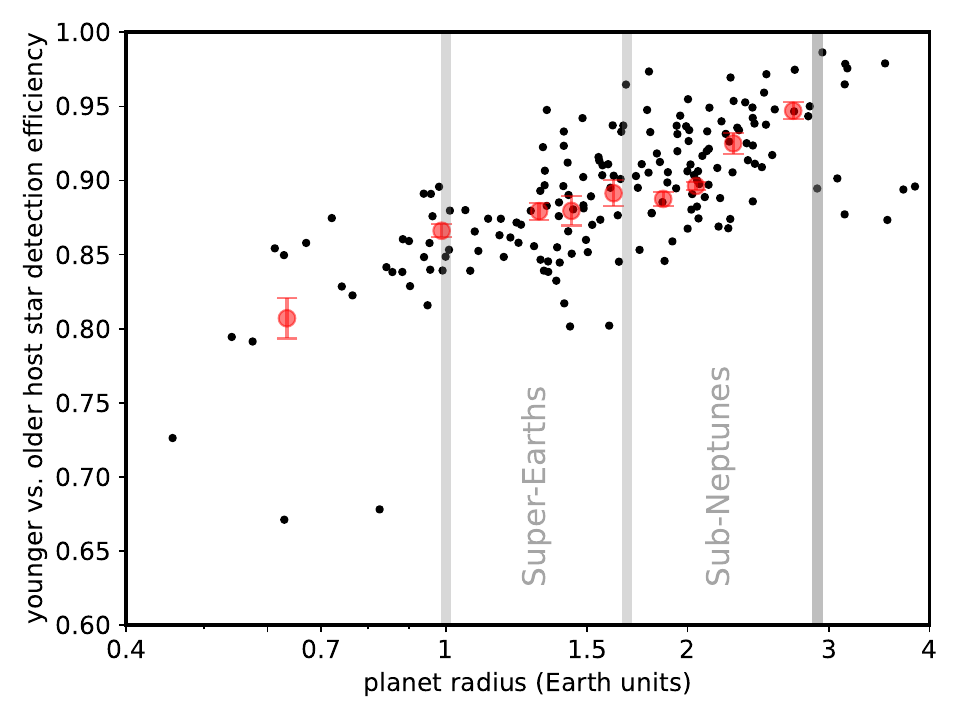}
    \caption{Ratio of mean planet detection efficiency between younger ($<$4 Gyr, based on rotation) and older sub-samples of the late K and early M dwarfs observed by \kepler{} with rotation periods reported in \citet{Santos2019}.  Each point represents a planet in our KOI sample, and the red points are binned averages of 20 planets with error bars representing the scatter within the bin divided by $\sqrt{20}$.  The approximate radius domains of super-Earths and sub-Neptunes are delineated.}
    \label{fig:detection}
\end{figure}

\emph{Planet detection bias:} We also investigated whether the observed radius evolution could be an artefact of differences in planet detection efficiency between young and old stars\footnote{The main sequence lifetime of stars increases with decreasing stellar mass but this is a consideration only for host stars more massive than the Sun.}.  Rotation-driven magnetic activity (spots and flaring) on cool stars introduces noise into lightcurves.  As stars age and spin down, the noise decreases \citep[e.g.,][]{Davenport2016}, meaning that, all else being equal, smaller and/or more distant planets can be detected around older stars.   To evaluate the magnitude of this systematic, we divided the \kepler{} late K/early M dwarf sample with rotation periods in \citet{Santos2019} into 773 (older) stars more slowly rotating than the rotation sequence of the 4 Gyr-old cluster M67 \citep{Dungee2022}, and 1936 (younger) stars that are more rapidly rotating than that sequence.  The age of M67 is within error of our median sample age (3.86 Gyr).  We calculated overall detection efficiency among the two samples for each planet in our KOI list and plot the ratio vs. planet radius in Fig. \ref{fig:detection}.  All planets in the radius range of interest are less detectable by about 10\% around the more rapidly rotating (``younger") stars than around the more slowly rotating (``older") stars and the difference is largest among the smallest planets.  However, the relative decline in detection efficiency between super-Earths and sub-Neptunes is $<$10\% and cannot explain the factor of $\sim$2 differences between the younger and older radius distributions (Fig. \ref{fig:radius}).  Because we cannot adequately filter the overall sample for binary stars, which tend to be more rapidly rotating than single stars and also dilute any transit signals, we do not explicitly correct for this difference in detection efficiency.   

\begin{figure}
\centering
\includegraphics[width=\columnwidth]{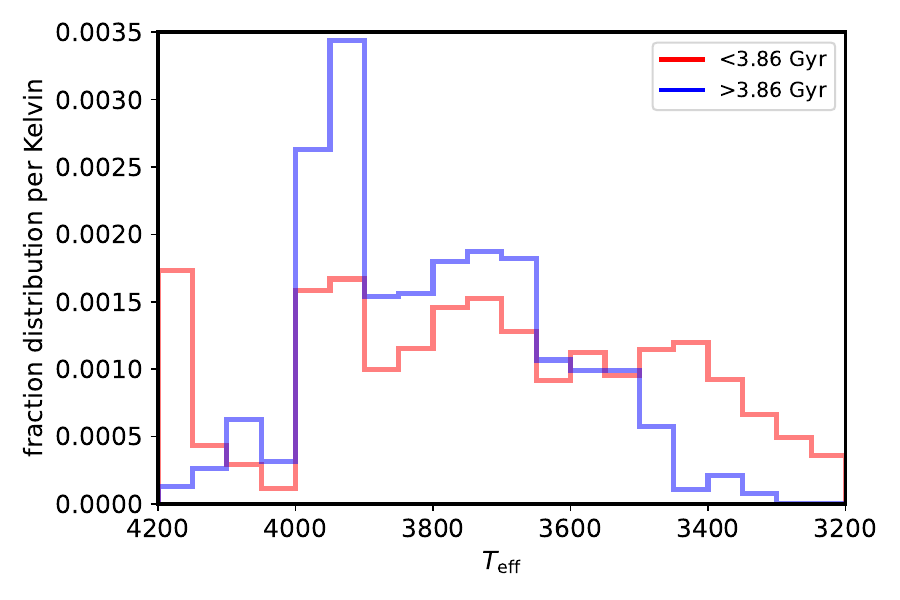}
\caption{Effective temperature distribution of younger ($<$4 Gyr, based on rotation) and older subsets of \kepler{} late K- and early M-type target stars, showing the difference due to selection effects (e.g., detectable rotation).}
\label{fig:teff}
\end{figure}

\begin{figure}
\centering
\includegraphics[width=\columnwidth]{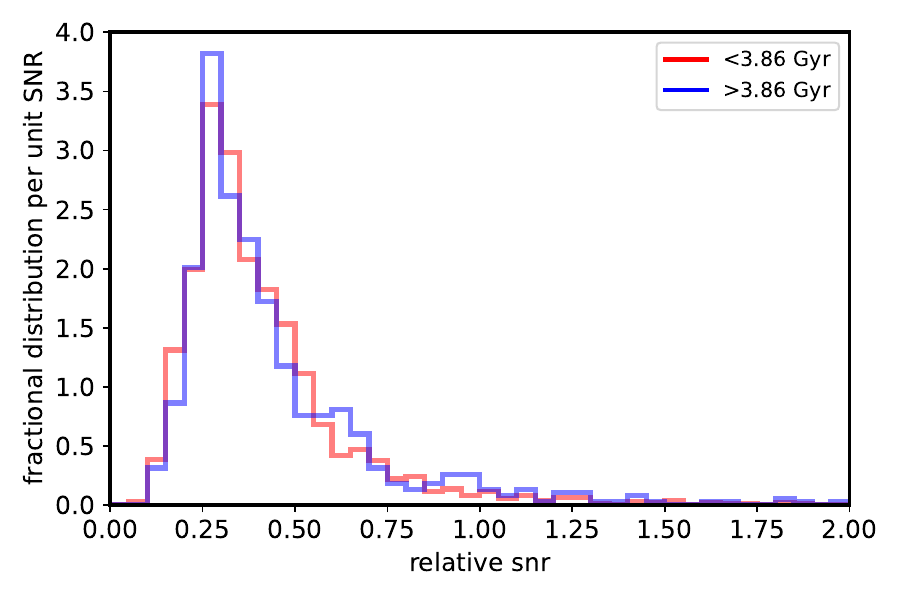}
\caption{Distribution of the relative detection SNR at a given planet radius and orbital period based on a scaling with stellar parameters (see text), showing no significant difference between younger and old stars.}
\label{fig:snr}
\end{figure}

\emph{Rotation period selection bias:} This observed change in planet radii could possibly arise from a difference in the properties of younger and older stars brought about by selection for detectable rotation periods.  Cooler dwarf stars tend to rotate more slowly at a given age \citep[e.g.,][]{Curtis2019b} and their rotation is more difficult to detect due to the longer time baseline needed to capture variability as well as the lower amplitude of variability from decreased magnetic activity.  Thus, cooler stars are more likely to be included in the younger sub-sample, an effect which is apparent in the \teff{} distribution of \kepler{} target stars (not just KOIs) with rotation-based ages younger or older than 3.86 Gyr (Fig. \ref{fig:teff}).  This bias with \teff{} should also bias the distributions with stellar radius, and potentially detection of transiting planets.  However, we find that this does not produce any significant difference in planet detectability.  The distribution of predicted transit signal-to-noise ratio (SNR) of the two subsets of target stars are nearly indistinguishable (Fig. \ref{fig:snr}).  The median values are 0.35 and 0.36 and a two-sided Kolmogorov-Smirnov (K-S) returns a probability of 0.13 that the two samples, excluding the tails with relative SNR $>$1.3, are drawn from the same distribution.  Here, we calculated the relative SNR for a given $R_p$ and $P_K$ as the inverse of the stellar radius times the square root of the brightness times the transit duration, which scales as $R_*^{-3/2}M_*^{-1/2}10^{-m/5}$, where \gaia{} $G$ magnitude was used as a proxy for \kepler{} magnitude $m$.  Cooler dwarf stars are smaller but also intrinsically fainter, and these two opposing trends apparently cancel out in this field.

\begin{figure}
    \centering
    \includegraphics[width=\columnwidth]{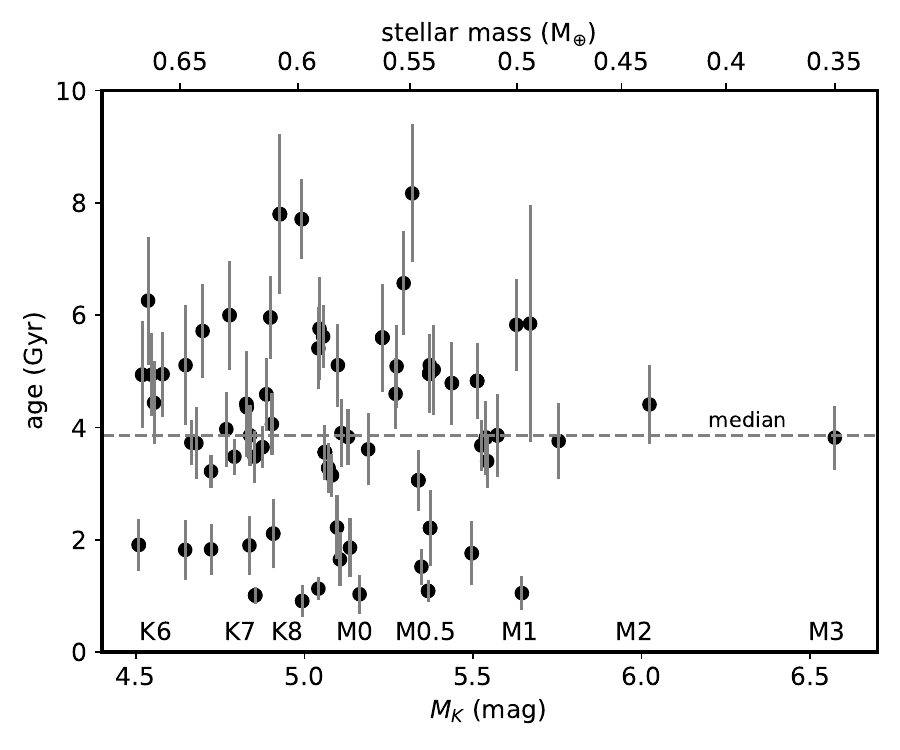}
    \caption{Distribution of gyrochronologic host star age vs. stellar luminosity (absolute $K_s$ -band magnitude), an age-independent proxy for mass for these main sequence stars.  Corresponding stellar masses and spectral types from \citet{Pecaut2013} are at the top and bottom axes, respectively.}
    \label{fig:age_mk}
\end{figure}

\emph{Planet population variation:}  A dependence of the intrinsic planet population on stellar mass, combined with a difference in the mass distribution between younger and older sub-samples, could mimic evolution.  The occurrence ratio of sub-Neptunes to super-Earths to sub-Neptunes modestly decreases with stellar mass between 0.5 and 1.4\msun{} \citep[about 0.04 per 0.1 dex in mass, top panels of Fig. 12 in ][]{Petigura2022}, but seems to markedly \emph{increase} with increasing stellar mass in the M dwarf range \citep{Hardegree-Ullman2019,Ment2023}.  Thus, if there was a bias towards more higher-mass stars in the older sample due to a detection bias for shorter rotation periods (see above), the older sample should contain \emph{more} sub-Neptunes, opposite to what is observed.  Moreover, there is no apparent trend between host star luminosity (here $M_K$, a proxy for mass) and age in our sample (Fig. \ref{fig:age_mk}).  A two-sided K-S test returns a probability of 0.29 that the luminosities of younger and older stars are drawn from the same distribution.    

\begin{figure}
    \centering
    \includegraphics[width=\columnwidth]{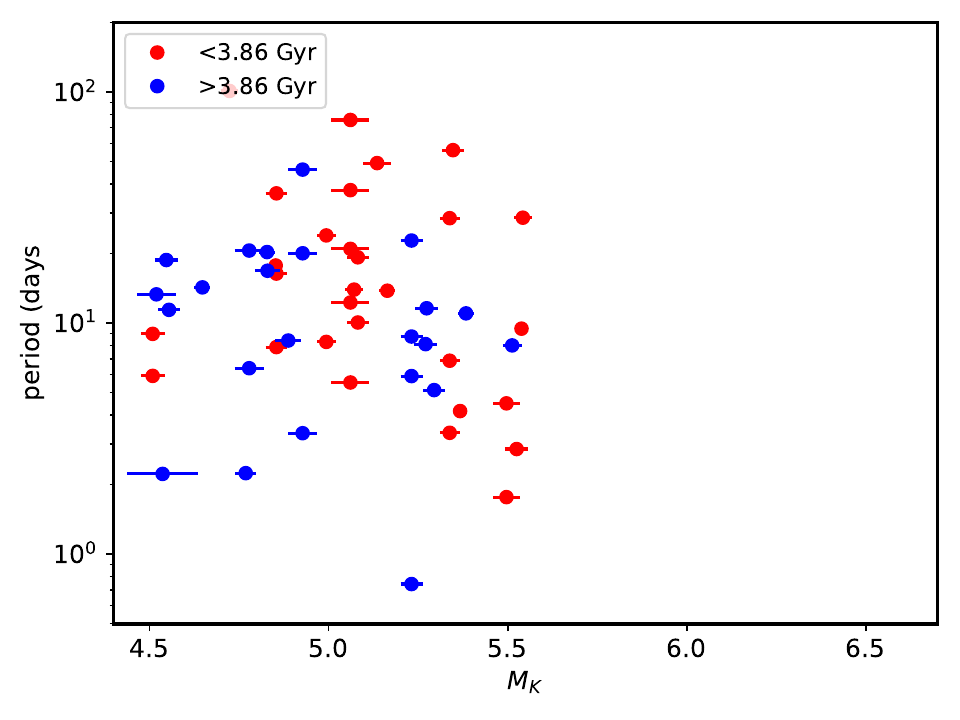}
    \caption{Orbital period vs. host star luminosity ($M_K$) for younger (red) and older (blue) sub-Neptunes (planets with radii between 1.7 and 4\rearth).  There is no correlation that could explain the absence of older sub-Neptunes at larger radii.}
    \label{fig:period-mk}
\end{figure}

% Don't change these lines
\bsp	% typesetting comment
\label{lastpage}
\end{document}

%% file: newcommands.tex
%\newcommand{\earth}{\mbox{$\oplus$}}

% Make Orcid icon
%\definecolor{lime}{HTML}{A6CE39}
\DeclareRobustCommand{\orcidicon}{%
    \begin{tikzpicture}
    \draw[lime, fill=lime] (0,0) 
    circle [radius=0.16] 
    node[white] {{\fontfamily{qag}\selectfont \tiny ID}};
    \draw[white, fill=white] (-0.0625,0.095) 
    circle [radius=0.007];
    \end{tikzpicture}
    \hspace{-2mm}
}

\newcommand*{\tabhead}[1]{\multicolumn{1}{c}{\bfseries #1}}

\newcommand{\mystar}{{\Large {\fontfamily{lmr}\selectfont$\star$}}}

% Journals
\newcommand{\jgrp}{J. Geophys. Res.-Planets}
\newcommand{\rsos}{Royal Soc. Open Sci.}
\newcommand{\astrolett}{Astron. Lett.}

\newcommand{\usco}{Upper~Sco}
\newcommand{\rhooph}{$\rho$~Oph}
\newcommand{\oi}{[O\,I]}
\newcommand{\suii}{[S\,II]}
\newcommand{\nii}{[N\,II]}
\newcommand{\nai}{Na\,I}
\newcommand{\caii}{Ca\,II}
\newcommand{\mdotacc}{$\cdot{M}_{\rm acc}$}
\newcommand{\mdotwind}{$\cdot{M}_{\rm wind}$}
\newcommand{\hei}{He\,I}
\newcommand{\lyalpha}{Ly-$\alpha$}
\newcommand{\halpha}{H$\alpha$}
\newcommand{\mgi}{Mg\,I}
\newcommand{\sii}{Si\,I}

% Missions

\newcommand{\rosat}{\emph{Rosat}}
\newcommand{\galex}{\emph{GALEX}}
\newcommand{\tess}{\emph{TESS}}
\newcommand{\plato}{\emph{PLATO}}
\newcommand{\gaia}{\emph{Gaia}}
\newcommand{\ktwo}{\emph{K2}}
\newcommand{\jwst}{\emph{JWST}}
\newcommand{\kepler}{\emph{Kepler}}
\newcommand{\corot}{\emph{CoRoT}}
\newcommand{\hipp}{\emph{Hipparcos}}
\newcommand{\spitzer}{\emph{Spizter}}
\newcommand{\herschel}{\emph{Herschel}}
\newcommand{\hst}{\emph{HST}}
\newcommand{\wise}{\emph{WISE}}
\newcommand{\swift}{\emph{Swift}}
\newcommand{\chandra}{\emph{Chandra}}

% astronomical objects
\newcommand{\twa}{TW Hydra}
\newcommand{\bpic}{$\beta$~Pictoris}
\newcommand{\abdor}{AB~Doradus}
\newcommand{\rup}{Ruprecht~147}
\newcommand{\etacha}{$\eta$\,Chamaeleontis}

\newcommand{\doar}{DoAr\,25}
\newcommand{\epcha}{EP\,Cha}
\newcommand{\rylup}{RY\,Lup}
\newcommand{\hdtwofour}{HD\,240779}

% parameters
\newcommand{\sigrv}{$\sigma_{\rm RV}$}

% units
\newcommand{\etal}{\mbox{\rm et al.~}}
\newcommand{\ms}{\mbox{m\,s$^{-1}~$}}
\newcommand{\ks}{\mbox{km\,s$^{-1}~$}}
\newcommand{\kse}{\mbox{km\,s$^{-1}$}}
\newcommand{\mse}{\mbox{m\,s$^{-1}$}}
\newcommand{\msy}{\mbox{m\,s$^{-1}$\,yr$^{-1}~$}}
\newcommand{\msye}{\mbox{m\,s$^{-1}$\,yr$^{-1}$}}
\newcommand{\msun}{M$_{\odot}$}
\newcommand{\msune}{M$_{\odot}$}
\newcommand{\rsun}{R$_{\odot}$}
\newcommand{\lsun}{L$_{\odot}~$}
\newcommand{\rsune}{R$_{\odot}$}
\newcommand{\mjup}{M$_{\rm JUP}~$}
\newcommand{\mjupe}{M$_{\rm JUP}$}
\newcommand{\msat}{M$_{\rm SAT}~$}
\newcommand{\msate}{M$_{\rm SAT}$}
\newcommand{\mnep}{M$_{\rm NEP}~$}
\newcommand{\mnepe}{M$_{\rm NEP}$}
\newcommand{\mearth}{$M_{\oplus}$}
\newcommand{\mearthe}{$M_{\oplus}$}
\newcommand{\rearth}{$R_{\oplus}$}
\newcommand{\rearthe}{$R_{\oplus}$}
\newcommand{\rjup}{R$_{\rm JUP}~$}
\newcommand{\msinie}{$M_{\rm p} \sin i$}
\newcommand{\vsinie}{$V \sin i$}
\newcommand{\mbsini}{$M_b \sin i~$}
\newcommand{\mcsini}{$M_c \sin i~$}
\newcommand{\mdsini}{$M_d \sin i~$}
\newcommand{\chisq}{$\chi_{\nu}^2$}
\newcommand{\chinu}{$\chi_{\nu}$}
\newcommand{\chinusq}{$\chi_{\nu}^2$}
\newcommand{\arel}{$a_{\rm rel}$}
\newcommand{\feh}{\ensuremath{[\mbox{Fe}/\mbox{H}]}}
\newcommand{\rphk}{\ensuremath{R'_{\mbox{\scriptsize HK}}}}
\newcommand{\lrphk}{\ensuremath{\log{\rphk}}}
\newcommand{\cs}{$\sqrt{\chi^2_{\nu}}$}
\newcommand{\etaearth}{$\mathbf \eta_{\oplus} ~$}
\newcommand{\etaearthe}{$\mathbf \eta_{\oplus}$}
\newcommand{\searth}{$S_{\oplus}$}

\newcommand{\sini}{\ensuremath{\sin i}}
\newcommand{\msini}{\ensuremath{M_{\rm p} \sin i}}
\newcommand{\mplsini}{\ensuremath{\mpl\sin i}}
\newcommand{\teff}{\ensuremath{T_{\rm eff}}}
\newcommand{\teq}{\ensuremath{T_{\rm eq}}}
\newcommand{\logg}{\ensuremath{\log{g}}}
\newcommand{\vsini}{\ensuremath{v \sin{i}}}
\newcommand{\ebv}{E($B$-$V$)}

% -------- Aliases specific to Kepler or Kepler diagnostics -------------
\newcommand{\Kepler}{\emph{Kepler}~}
\newcommand{\Keplere}{\emph{Kepler}}
\newcommand{\blender}{{\tt BLENDER}~}

% -------- Aliases about the star properties  -------------
%\newcommand{\kp}{\ensuremath{K_{p}}}
\newcommand{\kp}{\ensuremath{\mathrm{Kp}}}
\newcommand{\rstar}{\ensuremath{R_\star}}
\newcommand{\mstar}{\ensuremath{M_\star}}
\newcommand{\loggstar}{\ensuremath{\logg_\star}}
\newcommand{\lstar}{\ensuremath{L_\star}}
\newcommand{\astar}{\ensuremath{a_\star}}
\newcommand{\loglstar}{\ensuremath{\log{L_\star}}}
\newcommand{\rhostar}{\ensuremath{\rho_\star}}

% -------- Aliases about planet properties -------------
\newcommand{\rp}{\ensuremath{R_{\rm p}}}
\newcommand{\rpl}{\ensuremath{r_{\rm P}}~}
\newcommand{\rple}{\ensuremath{r_{\rm P}}}
\newcommand{\mpl}{\ensuremath{m_{\rm P}}~}
\newcommand{\lpl}{\ensuremath{L_{\rm P}}~}
\newcommand{\rhopl}{\ensuremath{\rho_{\rm P}}~}
\newcommand{\loggpl}{\ensuremath{\logg_{\rm P}}~}
\newcommand{\logrpl}{\ensuremath{\log r_{\rm P}}~}
\newcommand{\logrple}{\ensuremath{\log r_{\rm P}}}
\newcommand{\loga}{\ensuremath{\log a}~}
\newcommand{\logmpl}{\ensuremath{\log m_{\rm P}}~}

\newcommand{\water}{H$_2$O}
\newcommand{\methane}{CH$_4$}

%% file: table1.tex
\begin{tabular}[t]{lllll}
\hline
\multicolumn{1}{c}{Name} & \multicolumn{1}c{}{Disposition} & \multicolumn{1}{c}{Period} & \multicolumn{1}{c}{Radius} & \multicolumn{1}{c}{Age}\\
 &  & \multicolumn{1}{c}{days} & \multicolumn{1}{c}{\rearth} & \multicolumn{1}{c}{Gyr}\\
\hline
KOI-247.01 & confirmed & 13.82 & 1.91 +0.10/-0.11 & 1.03$\pm$0.35\\
KOI-251.01 & confirmed & 4.16 & 2.65 +0.10/-0.10 & 1.09$\pm$0.20\\
KOI-251.02 & confirmed & 5.77 & 0.85 +0.05/-0.05 & 1.09$\pm$0.20\\
KOI-253.01 & confirmed & 6.38 & 3.20 +0.13/-0.13 & 6.00$\pm$0.97\\
KOI-253.02 & confirmed & 20.62 & 1.85 +0.10/-0.11 & 6.00$\pm$0.97\\
KOI-254.01 & confirmed & 2.46 & 11.54 +0.35/-0.35 & 1.14$\pm$0.24\\
KOI-314.01 & confirmed & 13.78 & 1.50 +0.05/-0.05 & 2.21$\pm$0.67\\
KOI-314.02 & confirmed & 23.09 & 1.33 +0.05/-0.05 & 2.21$\pm$0.67\\
KOI-314.03 & confirmed & 10.31 & 0.64 +0.03/-0.03 & 2.21$\pm$0.67\\
KOI-430.01 & confirmed & 12.38 & 2.69 +0.10/-0.10 & ---\\
\hline
\end{tabular}

%% file: manuscript.bbl
\begin{thebibliography}{}
\makeatletter
\relax
\def\mn@urlcharsother{\let\do\@makeother \do\$\do\&\do\#\do\^\do\_\do\%\do\~}
\def\mn@doi{\begingroup\mn@urlcharsother \@ifnextchar [ {\mn@doi@} {\mn@doi@[]}}
\def\mn@doi@[#1]#2{\def\@tempa{#1}\ifx\@tempa\@empty \href {http://dx.doi.org/#2} {doi:#2}\else \href {http://dx.doi.org/#2} {#1}\fi \endgroup}
\def\mn@eprint#1#2{\mn@eprint@#1:#2::\@nil}
\def\mn@eprint@arXiv#1{\href {http://arxiv.org/abs/#1} {{\tt arXiv:#1}}}
\def\mn@eprint@dblp#1{\href {http://dblp.uni-trier.de/rec/bibtex/#1.xml} {dblp:#1}}
\def\mn@eprint@#1:#2:#3:#4\@nil{\def\@tempa {#1}\def\@tempb {#2}\def\@tempc {#3}\ifx \@tempc \@empty \let \@tempc \@tempb \let \@tempb \@tempa \fi \ifx \@tempb \@empty \def\@tempb {arXiv}\fi \@ifundefined {mn@eprint@\@tempb}{\@tempb:\@tempc}{\expandafter \expandafter \csname mn@eprint@\@tempb\endcsname \expandafter{\@tempc}}}

\bibitem[\protect\citeauthoryear{{Adams}, {Ciardi}, {Dupree}, {Gautier}, {Kulesa}  \& {McCarthy}}{{Adams} et~al.}{2012}]{Adams2012}
{Adams} E.~R.,  {Ciardi} D.~R.,  {Dupree} A.~K.,  {Gautier} T.~N. I.,  {Kulesa} C.,   {McCarthy} D.,  2012, \mn@doi [\aj] {10.1088/0004-6256/144/2/42}, \href {https://ui.adsabs.harvard.edu/abs/2012AJ....144...42A} {144, 42}

\bibitem[\protect\citeauthoryear{{Albrecht}, {Dawson}  \& {Winn}}{{Albrecht} et~al.}{2022}]{Albrecht2022}
{Albrecht} S.~H.,  {Dawson} R.~I.,   {Winn} J.~N.,  2022, \mn@doi [\pasp] {10.1088/1538-3873/ac6c09}, \href {https://ui.adsabs.harvard.edu/abs/2022PASP..134h2001A} {134, 082001}

\bibitem[\protect\citeauthoryear{{Barnes}, {Jackson}, {Raymond}, {West}  \& {Greenberg}}{{Barnes} et~al.}{2009}]{Barnes2009}
{Barnes} R.,  {Jackson} B.,  {Raymond} S.~N.,  {West} A.~A.,   {Greenberg} R.,  2009, \mn@doi [\apj] {10.1088/0004-637X/695/2/1006}, \href {http://adsabs.harvard.edu/abs/2009ApJ...695.1006B} {695, 1006}

\bibitem[\protect\citeauthoryear{{Batalha} et~al.,}{{Batalha} et~al.}{2013}]{Batalha2013}
{Batalha} N.~M.,  et~al., 2013, \mn@doi [\apjs] {10.1088/0067-0049/204/2/24}, \href {https://ui.adsabs.harvard.edu/abs/2013ApJS..204...24B} {204, 24}

\bibitem[\protect\citeauthoryear{{Belokurov} et~al.,}{{Belokurov} et~al.}{2020}]{Belokurov2020}
{Belokurov} V.,  et~al., 2020, \mn@doi [\mnras] {10.1093/mnras/staa1522}, \href {https://ui.adsabs.harvard.edu/abs/2020MNRAS.496.1922B} {496, 1922}

\bibitem[\protect\citeauthoryear{{Berger}, {Howard}  \& {Boesgaard}}{{Berger} et~al.}{2018}]{Berger2018b}
{Berger} T.~A.,  {Howard} A.~W.,   {Boesgaard} A.~M.,  2018, \mn@doi [\apj] {10.3847/1538-4357/aab154}, \href {http://adsabs.harvard.edu/abs/2018ApJ...855..115B} {855, 115}

\bibitem[\protect\citeauthoryear{{Berger}, {Huber}, {van Saders}, {Gaidos}, {Tayar}  \& {Kraus}}{{Berger} et~al.}{2020}]{Berger2020}
{Berger} T.~A.,  {Huber} D.,  {van Saders} J.~L.,  {Gaidos} E.,  {Tayar} J.,   {Kraus} A.~L.,  2020, \mn@doi [\aj] {10.3847/1538-3881/159/6/280}, \href {https://ui.adsabs.harvard.edu/abs/2020AJ....159..280B} {159, 280}

\bibitem[\protect\citeauthoryear{{B{\'e}thune} \& {Rafikov}}{{B{\'e}thune} \& {Rafikov}}{2019}]{William2019}
{B{\'e}thune} W.,  {Rafikov} R.~R.,  2019, \mn@doi [\mnras] {10.1093/mnras/stz1870}, \href {https://ui.adsabs.harvard.edu/abs/2019MNRAS.488.2365B} {488, 2365}

\bibitem[\protect\citeauthoryear{{Bonfanti} et~al.,}{{Bonfanti} et~al.}{2024}]{Bonfanti2024}
{Bonfanti} A.,  et~al., 2024, \mn@doi [\aap] {10.1051/0004-6361/202348180}, \href {https://ui.adsabs.harvard.edu/abs/2024A&A...682A..66B} {682, A66}

\bibitem[\protect\citeauthoryear{{Borucki} et~al.,}{{Borucki} et~al.}{2010}]{Borucki2010}
{Borucki} W.~J.,  et~al., 2010, \mn@doi [Science] {10.1126/science.1185402}, \href {https://ui.adsabs.harvard.edu/abs/2010Sci...327..977B} {327, 977}

\bibitem[\protect\citeauthoryear{{Brown}, {Latham}, {Everett}  \& {Esquerdo}}{{Brown} et~al.}{2011}]{Brown2011}
{Brown} T.~M.,  {Latham} D.~W.,  {Everett} M.~E.,   {Esquerdo} G.~A.,  2011, \mn@doi [\aj] {10.1088/0004-6256/142/4/112}, \href {http://adsabs.harvard.edu/abs/2011AJ....142..112B} {142, 112}

\bibitem[\protect\citeauthoryear{{Bryson} et~al.,}{{Bryson} et~al.}{2021}]{Bryson2021}
{Bryson} S.,  et~al., 2021, \mn@doi [\aj] {10.3847/1538-3881/abc418}, \href {https://ui.adsabs.harvard.edu/abs/2021AJ....161...36B} {161, 36}

\bibitem[\protect\citeauthoryear{{Burke} \& {Catanzarite}}{{Burke} \& {Catanzarite}}{2017a}]{Burke2017b}
{Burke} C.~J.,  {Catanzarite} J.,  2017a, {Planet Detection Metrics: Window and One-Sigma Depth Functions for Data Release 25}, Kepler Science Document KSCI-19101-002, id. 14. Edited by Michael R. Haas and Natalie M. Batalha

\bibitem[\protect\citeauthoryear{{Burke} \& {Catanzarite}}{{Burke} \& {Catanzarite}}{2017b}]{Burke2017}
{Burke} C.~J.,  {Catanzarite} J.,  2017b, {Planet Detection Metrics: Per-Target Detection Contours for Data Release 25}, Kepler Science Document KSCI-19111-002, id. 19. Edited by Michael R. Haas and Natalie M. Batalha

\bibitem[\protect\citeauthoryear{{Burke} et~al.,}{{Burke} et~al.}{2015}]{Burke2015}
{Burke} C.~J.,  et~al., 2015, \mn@doi [\apj] {10.1088/0004-637X/809/1/8}, \href {https://ui.adsabs.harvard.edu/abs/2015ApJ...809....8B} {809, 8}

\bibitem[\protect\citeauthoryear{{Burn}, {Mordasini}, {Mishra}, {Haldemann}, {Venturini}, {Emsenhuber}  \& {Henning}}{{Burn} et~al.}{2024}]{Burn2024}
{Burn} R.,  {Mordasini} C.,  {Mishra} L.,  {Haldemann} J.,  {Venturini} J.,  {Emsenhuber} A.,   {Henning} T.,  2024, \mn@doi [Nature Astronomy] {10.1038/s41550-023-02183-7}, \href {https://ui.adsabs.harvard.edu/abs/2024NatAs...8..463B} {8, 463}

\bibitem[\protect\citeauthoryear{{Ca{\~n}as} et~al.,}{{Ca{\~n}as} et~al.}{2020}]{Canas2020}
{Ca{\~n}as} C.~I.,  et~al., 2020, \mn@doi [\aj] {10.3847/1538-3881/abac67}, \href {https://ui.adsabs.harvard.edu/abs/2020AJ....160..147C} {160, 147}

\bibitem[\protect\citeauthoryear{{Chen} et~al.,}{{Chen} et~al.}{2022}]{ChenD-C2022}
{Chen} D.-C.,  et~al., 2022, \mn@doi [\aj] {10.3847/1538-3881/ac641f}, \href {https://ui.adsabs.harvard.edu/abs/2022AJ....163..249C} {163, 249}

\bibitem[\protect\citeauthoryear{{Christiansen} et~al.,}{{Christiansen} et~al.}{2012}]{Christiansen2012}
{Christiansen} J.~L.,  et~al., 2012, \mn@doi [\pasp] {10.1086/668847}, \href {https://ui.adsabs.harvard.edu/abs/2012PASP..124.1279C} {124, 1279}

\bibitem[\protect\citeauthoryear{{Claytor}, {van Saders}, {Llama}, {Sadowski}, {Quach}  \& {Avallone}}{{Claytor} et~al.}{2022}]{Claytor2022}
{Claytor} Z.~R.,  {van Saders} J.~L.,  {Llama} J.,  {Sadowski} P.,  {Quach} B.,   {Avallone} E.~A.,  2022, \mn@doi [\apj] {10.3847/1538-4357/ac498f}, \href {https://ui.adsabs.harvard.edu/abs/2022ApJ...927..219C} {927, 219}

\bibitem[\protect\citeauthoryear{{Cloutier} \& {Menou}}{{Cloutier} \& {Menou}}{2020}]{Cloutier2020}
{Cloutier} R.,  {Menou} K.,  2020, \mn@doi [\aj] {10.3847/1538-3881/ab8237}, \href {https://ui.adsabs.harvard.edu/abs/2020AJ....159..211C} {159, 211}

\bibitem[\protect\citeauthoryear{{Curtis}, {Ag{\"u}eros}, {Douglas}  \& {Meibom}}{{Curtis} et~al.}{2019}]{Curtis2019b}
{Curtis} J.~L.,  {Ag{\"u}eros} M.~A.,  {Douglas} S.~T.,   {Meibom} S.,  2019, \mn@doi [\apj] {10.3847/1538-4357/ab2393}, \href {https://ui.adsabs.harvard.edu/abs/2019ApJ...879...49C} {879, 49}

\bibitem[\protect\citeauthoryear{{Cutri} et~al.,}{{Cutri} et~al.}{2013}]{Cutri2013}
{Cutri} R.~M.,  et~al., 2013, Technical report, {Explanatory Supplement to the AllWISE Data Release Products}.
IPAC/Caltech

\bibitem[\protect\citeauthoryear{{Davenport}}{{Davenport}}{2016}]{Davenport2016}
{Davenport} J. R.~A.,  2016, \mn@doi [\apj] {10.3847/0004-637X/829/1/23}, \href {https://ui.adsabs.harvard.edu/abs/2016ApJ...829...23D} {829, 23}

\bibitem[\protect\citeauthoryear{{David} et~al.,}{{David} et~al.}{2021}]{David2021}
{David} T.~J.,  et~al., 2021, \mn@doi [\aj] {10.3847/1538-3881/abf439}, \href {https://ui.adsabs.harvard.edu/abs/2021AJ....161..265D} {161, 265}

\bibitem[\protect\citeauthoryear{{Dressing} \& {Charbonneau}}{{Dressing} \& {Charbonneau}}{2015}]{Dressing2015}
{Dressing} C.~D.,  {Charbonneau} D.,  2015, \mn@doi [\apj] {10.1088/0004-637X/807/1/45}, \href {http://adsabs.harvard.edu/abs/2015ApJ...807...45D} {807, 45}

\bibitem[\protect\citeauthoryear{{Dungee}, {van Saders}, {Gaidos}, {Chun}, {Garc{\'\i}a}, {Magnier}, {Mathur}  \& {Santos}}{{Dungee} et~al.}{2022}]{Dungee2022}
{Dungee} R.,  {van Saders} J.,  {Gaidos} E.,  {Chun} M.,  {Garc{\'\i}a} R.~A.,  {Magnier} E.~A.,  {Mathur} S.,   {Santos} {\^A}. R.~G.,  2022, \mn@doi [\apj] {10.3847/1538-4357/ac90be}, \href {https://ui.adsabs.harvard.edu/abs/2022ApJ...938..118D} {938, 118}

\bibitem[\protect\citeauthoryear{{El-Badry} \& {Rix}}{{El-Badry} \& {Rix}}{2018}]{El-Badry2018}
{El-Badry} K.,  {Rix} H.-W.,  2018, \mn@doi [\mnras] {10.1093/mnras/sty2186}, \href {https://ui.adsabs.harvard.edu/abs/2018MNRAS.480.4884E} {480, 4884}

\bibitem[\protect\citeauthoryear{{El-Badry}, {Rix}  \& {Heintz}}{{El-Badry} et~al.}{2021}]{El-Badry2021}
{El-Badry} K.,  {Rix} H.-W.,   {Heintz} T.~M.,  2021, \mn@doi [\mnras] {10.1093/mnras/stab323}, \href {https://ui.adsabs.harvard.edu/abs/2021MNRAS.506.2269E} {506, 2269}

\bibitem[\protect\citeauthoryear{{France} et~al.,}{{France} et~al.}{2016}]{France2016}
{France} K.,  et~al., 2016, \mn@doi [\apj] {10.3847/0004-637X/820/2/89}, \href {http://adsabs.harvard.edu/abs/2016ApJ...820...89F} {820, 89}

\bibitem[\protect\citeauthoryear{{Fulton} et~al.,}{{Fulton} et~al.}{2017}]{Fulton2017}
{Fulton} B.~J.,  et~al., 2017, \mn@doi [\aj] {10.3847/1538-3881/aa80eb}, \href {http://adsabs.harvard.edu/abs/2017AJ....154..109F} {154, 109}

\bibitem[\protect\citeauthoryear{{Gaia Collaboration} et~al.,}{{Gaia Collaboration} et~al.}{2016}]{Gaia2016}
{Gaia Collaboration} et~al., 2016, \mn@doi [\aap] {10.1051/0004-6361/201629272}, \href {https://ui.adsabs.harvard.edu/abs/2016A&A...595A...1G} {595, A1}

\bibitem[\protect\citeauthoryear{{Gaia Collaboration} et~al.,}{{Gaia Collaboration} et~al.}{2023}]{Gaia2023}
{Gaia Collaboration} et~al., 2023, \mn@doi [\aap] {10.1051/0004-6361/202243940}, \href {https://ui.adsabs.harvard.edu/abs/2023A&A...674A...1G} {674, A1}

\bibitem[\protect\citeauthoryear{{Gaidos}}{{Gaidos}}{2017}]{Gaidos2017b}
{Gaidos} E.,  2017, \mn@doi [\mnras] {10.1093/mnrasl/slx063}, \href {http://adsabs.harvard.edu/abs/2017MNRAS.470L...1G} {470, L1}

\bibitem[\protect\citeauthoryear{{Gaidos}, {Mann}, {Kraus}  \& {Ireland}}{{Gaidos} et~al.}{2016}]{Gaidos2016}
{Gaidos} E.,  {Mann} A.~W.,  {Kraus} A.~L.,   {Ireland} M.,  2016, \mn@doi [\mnras] {10.1093/mnras/stw097}, 457, 2877

\bibitem[\protect\citeauthoryear{{Gaidos} et~al.,}{{Gaidos} et~al.}{2017}]{Gaidos2017a}
{Gaidos} E.,  et~al., 2017, \mn@doi [\mnras] {10.1093/mnras/stw2345}, 464, 850

\bibitem[\protect\citeauthoryear{{Gaidos}, {Claytor}, {Dungee}, {Ali}  \& {Feiden}}{{Gaidos} et~al.}{2023}]{Gaidos2023b}
{Gaidos} E.,  {Claytor} Z.,  {Dungee} R.,  {Ali} A.,   {Feiden} G.~A.,  2023, \mn@doi [\mnras] {10.1093/mnras/stad343}, \href {https://ui.adsabs.harvard.edu/abs/2023MNRAS.520.5283G} {520, 5283}

\bibitem[\protect\citeauthoryear{{Gill} et~al.,}{{Gill} et~al.}{2020}]{Gill2020}
{Gill} S.,  et~al., 2020, \mn@doi [\apjl] {10.3847/2041-8213/ab9eb9}, \href {https://ui.adsabs.harvard.edu/abs/2020ApJ...898L..11G} {898, L11}

\bibitem[\protect\citeauthoryear{{Greene}, {Bell}, {Ducrot}, {Dyrek}, {Lagage}  \& {Fortney}}{{Greene} et~al.}{2023}]{Greene2023}
{Greene} T.~P.,  {Bell} T.~J.,  {Ducrot} E.,  {Dyrek} A.,  {Lagage} P.-O.,   {Fortney} J.~J.,  2023, \mn@doi [\nat] {10.1038/s41586-023-05951-7}, \href {https://ui.adsabs.harvard.edu/abs/2023Natur.618...39G} {618, 39}

\bibitem[\protect\citeauthoryear{{Gupta} \& {Schlichting}}{{Gupta} \& {Schlichting}}{2019}]{Gupta2019}
{Gupta} A.,  {Schlichting} H.~E.,  2019, \mn@doi [\mnras] {10.1093/mnras/stz1230}, \href {https://ui.adsabs.harvard.edu/abs/2019MNRAS.487...24G} {487, 24}

\bibitem[\protect\citeauthoryear{{Gupta} \& {Schlichting}}{{Gupta} \& {Schlichting}}{2021}]{Gupta2021}
{Gupta} A.,  {Schlichting} H.~E.,  2021, \mn@doi [\mnras] {10.1093/mnras/stab1128}, \href {https://ui.adsabs.harvard.edu/abs/2021MNRAS.504.4634G} {504, 4634}

\bibitem[\protect\citeauthoryear{{Hardegree-Ullman}, {Cushing}, {Muirhead}  \& {Christiansen}}{{Hardegree-Ullman} et~al.}{2019}]{Hardegree-Ullman2019}
{Hardegree-Ullman} K.~K.,  {Cushing} M.~C.,  {Muirhead} P.~S.,   {Christiansen} J.~L.,  2019, \mn@doi [\aj] {10.3847/1538-3881/ab21d2}, \href {https://ui.adsabs.harvard.edu/abs/2019AJ....158...75H} {158, 75}

\bibitem[\protect\citeauthoryear{{Hirano} et~al.,}{{Hirano} et~al.}{2018}]{Hirano2018}
{Hirano} T.,  et~al., 2018, \mn@doi [\aj] {10.3847/1538-3881/aaa9c1}, \href {http://adsabs.harvard.edu/abs/2018AJ....155..127H} {155, 127}

\bibitem[\protect\citeauthoryear{{Ho} \& {Van Eylen}}{{Ho} \& {Van Eylen}}{2023}]{Ho2023}
{Ho} C. S.~K.,  {Van Eylen} V.,  2023, \mn@doi [\mnras] {10.1093/mnras/stac3802}, \href {https://ui.adsabs.harvard.edu/abs/2023MNRAS.519.4056H} {519, 4056}

\bibitem[\protect\citeauthoryear{{Ho}, {Rogers}, {Van Eylen}, {Owen}  \& {Schlichting}}{{Ho} et~al.}{2024}]{Ho2024}
{Ho} C. S.~K.,  {Rogers} J.~G.,  {Van Eylen} V.,  {Owen} J.~E.,   {Schlichting} H.~E.,  2024, \mn@doi [\mnras] {10.1093/mnras/stae1376}, \href {https://ui.adsabs.harvard.edu/abs/2024MNRAS.531.3698H} {531, 3698}

\bibitem[\protect\citeauthoryear{{Hoffman} \& {Rowe}}{{Hoffman} \& {Rowe}}{2017}]{Hoffman2017}
{Hoffman} Kelsey L.,  {Rowe} J.~F.,  2017, {Uniform Modeling of KOIs: MCMC Notes for Data Release 25}, Kepler Science Document KSCI-19113-001, id. 21. Edited by Michael R. Haas and Natalie M. Batalha

\bibitem[\protect\citeauthoryear{{Holmberg} \& {Madhusudhan}}{{Holmberg} \& {Madhusudhan}}{2024}]{Holmberg2024}
{Holmberg} M.,  {Madhusudhan} N.,  2024, \mn@doi [\aap] {10.1051/0004-6361/202348238}, \href {https://ui.adsabs.harvard.edu/abs/2024A&A...683L...2H} {683, L2}

\bibitem[\protect\citeauthoryear{{Howard} et~al.,}{{Howard} et~al.}{2012}]{Howard2012}
{Howard} A.~W.,  et~al., 2012, \mn@doi [\apjs] {10.1088/0067-0049/201/2/15}, \href {http://adsabs.harvard.edu/abs/2012ApJS..201...15H} {201, 15}

\bibitem[\protect\citeauthoryear{{Howe} \& {Burrows}}{{Howe} \& {Burrows}}{2015}]{Howe2015}
{Howe} A.~R.,  {Burrows} A.,  2015, \mn@doi [\apj] {10.1088/0004-637X/808/2/150}, \href {http://adsabs.harvard.edu/abs/2015ApJ...808..150H} {808, 150}

\bibitem[\protect\citeauthoryear{{Hsu}, {Ford}, {Ragozzine}  \& {Ashby}}{{Hsu} et~al.}{2019}]{Hsu2019}
{Hsu} D.~C.,  {Ford} E.~B.,  {Ragozzine} D.,   {Ashby} K.,  2019, \mn@doi [\aj] {10.3847/1538-3881/ab31ab}, \href {https://ui.adsabs.harvard.edu/abs/2019AJ....158..109H} {158, 109}

\bibitem[\protect\citeauthoryear{{Innes}, {Tsai}  \& {Pierrehumbert}}{{Innes} et~al.}{2023}]{Innes2023}
{Innes} H.,  {Tsai} S.-M.,   {Pierrehumbert} R.~T.,  2023, \mn@doi [\apj] {10.3847/1538-4357/ace346}, \href {https://ui.adsabs.harvard.edu/abs/2023ApJ...953..168I} {953, 168}

\bibitem[\protect\citeauthoryear{{Johnson} et~al.,}{{Johnson} et~al.}{2012}]{Johnson2012b}
{Johnson} J.~A.,  et~al., 2012, \mn@doi [\aj] {10.1088/0004-6256/143/5/111}, \href {https://ui.adsabs.harvard.edu/abs/2012AJ....143..111J} {143, 111}

\bibitem[\protect\citeauthoryear{{Johnstone}, {Bartel}  \& {G{\"u}del}}{{Johnstone} et~al.}{2021}]{Johnstone2021}
{Johnstone} C.~P.,  {Bartel} M.,   {G{\"u}del} M.,  2021, \mn@doi [\aap] {10.1051/0004-6361/202038407}, \href {https://ui.adsabs.harvard.edu/abs/2021A&A...649A..96J} {649, A96}

\bibitem[\protect\citeauthoryear{{Kanodia}, {Wolfgang}, {Stefansson}, {Ning}  \& {Mahadevan}}{{Kanodia} et~al.}{2019}]{Kanodia2019}
{Kanodia} S.,  {Wolfgang} A.,  {Stefansson} G.~K.,  {Ning} B.,   {Mahadevan} S.,  2019, \mn@doi [\apj] {10.3847/1538-4357/ab334c}, \href {https://ui.adsabs.harvard.edu/abs/2019ApJ...882...38K} {882, 38}

\bibitem[\protect\citeauthoryear{{Kim}, {Wei}, {Chariton}, {Prakapenka}, {Ryu}, {Yang}  \& {Shim}}{{Kim} et~al.}{2023}]{Kim2023}
{Kim} T.,  {Wei} X.,  {Chariton} S.,  {Prakapenka} V.~B.,  {Ryu} Y.-J.,  {Yang} S.,   {Shim} S.-H.,  2023, \mn@doi [Proceedings of the National Academy of Science] {10.1073/pnas.2309786120}, \href {https://ui.adsabs.harvard.edu/abs/2023PNAS..12009786K} {120, e2309786120}

\bibitem[\protect\citeauthoryear{{King} \& {Wheatley}}{{King} \& {Wheatley}}{2021}]{King2021}
{King} G.~W.,  {Wheatley} P.~J.,  2021, \mn@doi [\mnras] {10.1093/mnrasl/slaa186}, \href {https://ui.adsabs.harvard.edu/abs/2021MNRAS.501L..28K} {501, L28}

\bibitem[\protect\citeauthoryear{{Kite}, {Fegley}, {Schaefer}  \& {Ford}}{{Kite} et~al.}{2020}]{Kite2020}
{Kite} E.~S.,  {Fegley} Bruce J.,  {Schaefer} L.,   {Ford} E.~B.,  2020, \mn@doi [\apj] {10.3847/1538-4357/ab6ffb}, \href {https://ui.adsabs.harvard.edu/abs/2020ApJ...891..111K} {891, 111}

\bibitem[\protect\citeauthoryear{{Kodama}, {Nitta}, {Genda}, {Takao}, {O'ishi}, {Abe-Ouchi}  \& {Abe}}{{Kodama} et~al.}{2018}]{Kodama2018}
{Kodama} T.,  {Nitta} A.,  {Genda} H.,  {Takao} Y.,  {O'ishi} R.,  {Abe-Ouchi} A.,   {Abe} Y.,  2018, \mn@doi [Journal of Geophysical Research (Planets)] {10.1002/2017JE005383}, \href {https://ui.adsabs.harvard.edu/abs/2018JGRE..123..559K} {123, 559}

\bibitem[\protect\citeauthoryear{{Kopparapu} et~al.,}{{Kopparapu} et~al.}{2013}]{Kopparapu2013}
{Kopparapu} R.~K.,  et~al., 2013, \mn@doi [\apj] {10.1088/0004-637X/765/2/131}, \href {http://adsabs.harvard.edu/abs/2013ApJ...765..131K} {765, 131}

\bibitem[\protect\citeauthoryear{{Kraus}, {Ireland}, {Huber}, {Mann}  \& {Dupuy}}{{Kraus} et~al.}{2016}]{Kraus2016}
{Kraus} A.~L.,  {Ireland} M.~J.,  {Huber} D.,  {Mann} A.~W.,   {Dupuy} T.~J.,  2016, \mn@doi [\aj] {10.3847/0004-6256/152/1/8}, \href {http://adsabs.harvard.edu/abs/2016AJ....152....8K} {152, 8}

\bibitem[\protect\citeauthoryear{{Leconte}, {Forget}, {Charnay}, {Wordsworth}  \& {Pottier}}{{Leconte} et~al.}{2013}]{Leconte2013}
{Leconte} J.,  {Forget} F.,  {Charnay} B.,  {Wordsworth} R.,   {Pottier} A.,  2013, \mn@doi [\nat] {10.1038/nature12827}, \href {https://ui.adsabs.harvard.edu/abs/2013Natur.504..268L} {504, 268}

\bibitem[\protect\citeauthoryear{{Lindegren} et~al.,}{{Lindegren} et~al.}{2018}]{Lindegren2018}
{Lindegren} L.,  et~al., 2018, \mn@doi [\aap] {10.1051/0004-6361/201832727}, \href {https://ui.adsabs.harvard.edu/abs/2018A&A...616A...2L} {616, A2}

\bibitem[\protect\citeauthoryear{{Lindegren} et~al.,}{{Lindegren} et~al.}{2021}]{Lindegren2021}
{Lindegren} L.,  et~al., 2021, \mn@doi [\aap] {10.1051/0004-6361/202039709}, \href {https://ui.adsabs.harvard.edu/abs/2021A&A...649A...2L} {649, A2}

\bibitem[\protect\citeauthoryear{{Lopez} \& {Rice}}{{Lopez} \& {Rice}}{2018}]{Lopez2018}
{Lopez} E.~D.,  {Rice} K.,  2018, \mn@doi [\mnras] {10.1093/mnras/sty1707}, \href {https://ui.adsabs.harvard.edu/abs/2018MNRAS.479.5303L} {479, 5303}

\bibitem[\protect\citeauthoryear{{Luger} \& {Barnes}}{{Luger} \& {Barnes}}{2015}]{Luger2015}
{Luger} R.,  {Barnes} R.,  2015, \mn@doi [Astrobiology] {10.1089/ast.2014.1231}, \href {https://ui.adsabs.harvard.edu/abs/2015AsBio..15..119L} {15, 119}

\bibitem[\protect\citeauthoryear{{Luque} \& {Pall{\'e}}}{{Luque} \& {Pall{\'e}}}{2022}]{Luque2022}
{Luque} R.,  {Pall{\'e}} E.,  2022, \mn@doi [Science] {10.1126/science.abl7164}, \href {https://ui.adsabs.harvard.edu/abs/2022Sci...377.1211L} {377, 1211}

\bibitem[\protect\citeauthoryear{{Magliano} et~al.,}{{Magliano} et~al.}{2024}]{Magliano2024}
{Magliano} C.,  et~al., 2024, \mn@doi [\mnras] {10.1093/mnras/stae210}, \href {https://ui.adsabs.harvard.edu/abs/2024MNRAS.528.2851M} {528, 2851}

\bibitem[\protect\citeauthoryear{{Mallama}, {Wang}  \& {Howard}}{{Mallama} et~al.}{2006}]{Mallama2006}
{Mallama} A.,  {Wang} D.,   {Howard} R.~A.,  2006, \mn@doi [\icarus] {10.1016/j.icarus.2005.12.014}, \href {https://ui.adsabs.harvard.edu/abs/2006Icar..182...10M} {182, 10}

\bibitem[\protect\citeauthoryear{{Mann}, {Gaidos}  \& {Ansdell}}{{Mann} et~al.}{2013}]{Mann2013}
{Mann} A.~W.,  {Gaidos} E.,   {Ansdell} M.,  2013, \mn@doi [\apj] {10.1088/0004-637X/779/2/188}, \href {http://adsabs.harvard.edu/abs/2013ApJ...779..188M} {779, 188}

\bibitem[\protect\citeauthoryear{{Mann}, {Feiden}, {Gaidos}, {Boyajian}  \& {von Braun}}{{Mann} et~al.}{2015}]{Mann2015}
{Mann} A.~W.,  {Feiden} G.~A.,  {Gaidos} E.,  {Boyajian} T.,   {von Braun} K.,  2015, \mn@doi [\apj] {10.1088/0004-637X/804/1/64}, \href {http://adsabs.harvard.edu/abs/2015ApJ...804...64M} {804, 64}

\bibitem[\protect\citeauthoryear{{Mann} et~al.,}{{Mann} et~al.}{2019}]{Mann2019}
{Mann} A.~W.,  et~al., 2019, \mn@doi [\apj] {10.3847/1538-4357/aaf3bc}, \href {https://ui.adsabs.harvard.edu/\#abs/2019ApJ...871...63M} {871, 63}

\bibitem[\protect\citeauthoryear{{Mathur} et~al.,}{{Mathur} et~al.}{2017}]{Mathur2017}
{Mathur} S.,  et~al., 2017, \mn@doi [\apjs] {10.3847/1538-4365/229/2/30}, \href {https://ui.adsabs.harvard.edu/abs/2017ApJS..229...30M} {229, 30}

\bibitem[\protect\citeauthoryear{{Mazeh}, {Holczer}  \& {Faigler}}{{Mazeh} et~al.}{2016}]{Mazeh2016}
{Mazeh} T.,  {Holczer} T.,   {Faigler} S.,  2016, \mn@doi [\aap] {10.1051/0004-6361/201528065}, \href {https://ui.adsabs.harvard.edu/abs/2016A&A...589A..75M} {589, A75}

\bibitem[\protect\citeauthoryear{{McDonald}, {Kreidberg}  \& {Lopez}}{{McDonald} et~al.}{2019}]{McDonald2019}
{McDonald} G.~D.,  {Kreidberg} L.,   {Lopez} E.,  2019, \mn@doi [\apj] {10.3847/1538-4357/ab1095}, \href {https://ui.adsabs.harvard.edu/abs/2019ApJ...876...22M} {876, 22}

\bibitem[\protect\citeauthoryear{{McLean} \& {Chaffee}}{{McLean} \& {Chaffee}}{2000}]{McLean2004}
{McLean} I.~S.,  {Chaffee} F.~H.,  2000, in {Iye} M.,  {Moorwood} A.~F.,  eds,  Society of Photo-Optical Instrumentation Engineers (SPIE) Conference Series Vol. 4008, Optical and IR Telescope Instrumentation and Detectors. pp~2--7, \mn@doi{10.1117/12.395404}

\bibitem[\protect\citeauthoryear{{Ment} \& {Charbonneau}}{{Ment} \& {Charbonneau}}{2023}]{Ment2023}
{Ment} K.,  {Charbonneau} D.,  2023, \mn@doi [\aj] {10.3847/1538-3881/acd175}, \href {https://ui.adsabs.harvard.edu/abs/2023AJ....165..265M} {165, 265}

\bibitem[\protect\citeauthoryear{{Messina} et~al.,}{{Messina} et~al.}{2017}]{Messina2017}
{Messina} S.,  et~al., 2017, \mn@doi [\aap] {10.1051/0004-6361/201730444}, \href {https://ui.adsabs.harvard.edu/abs/2017A&A...607A...3M} {607, A3}

\bibitem[\protect\citeauthoryear{{Moroz} et~al.,}{{Moroz} et~al.}{1985}]{Moroz1985}
{Moroz} V.~I.,  et~al., 1985, \mn@doi [Advances in Space Research] {10.1016/0273-1177(85)90202-9}, \href {https://ui.adsabs.harvard.edu/abs/1985AdSpR...5k.197M} {5, 197}

\bibitem[\protect\citeauthoryear{{Morton}}{{Morton}}{2015}]{Morton2015}
{Morton} T.~D.,  2015, {VESPA: False positive probabilities calculator}, Astrophysics Source Code Library (\mn@eprint {ascl} {1503.011})

\bibitem[\protect\citeauthoryear{{Morton}, {Bryson}, {Coughlin}, {Rowe}, {Ravichandran}, {Petigura}, {Haas}  \& {Batalha}}{{Morton} et~al.}{2016}]{Morton2016}
{Morton} T.~D.,  {Bryson} S.~T.,  {Coughlin} J.~L.,  {Rowe} J.~F.,  {Ravichandran} G.,  {Petigura} E.~A.,  {Haas} M.~R.,   {Batalha} N.~M.,  2016, \mn@doi [\apj] {10.3847/0004-637X/822/2/86}, \href {http://adsabs.harvard.edu/abs/2016ApJ...822...86M} {822, 86}

\bibitem[\protect\citeauthoryear{{Mulders}, {Pascucci}  \& {Apai}}{{Mulders} et~al.}{2015}]{Mulders2015}
{Mulders} G.~D.,  {Pascucci} I.,   {Apai} D.,  2015, \mn@doi [\apj] {10.1088/0004-637X/798/2/112}, \href {http://adsabs.harvard.edu/abs/2015ApJ...798..112M} {798, 112}

\bibitem[\protect\citeauthoryear{{Nascimbeni} et~al.,}{{Nascimbeni} et~al.}{2022}]{Nascimbeni2022}
{Nascimbeni} V.,  et~al., 2022, \mn@doi [\aap] {10.1051/0004-6361/202142256}, \href {https://ui.adsabs.harvard.edu/abs/2022A&A...658A..31N} {658, A31}

\bibitem[\protect\citeauthoryear{{Owen}}{{Owen}}{2019}]{Owen2019}
{Owen} J.~E.,  2019, \mn@doi [Annual Review of Earth and Planetary Sciences] {10.1146/annurev-earth-053018-060246}, \href {https://ui.adsabs.harvard.edu/abs/2019AREPS..47...67O} {47, 67}

\bibitem[\protect\citeauthoryear{{Owen} \& {Jackson}}{{Owen} \& {Jackson}}{2012}]{Owen2012}
{Owen} J.~E.,  {Jackson} A.~P.,  2012, \mn@doi [\mnras] {10.1111/j.1365-2966.2012.21481.x}, \href {https://ui.adsabs.harvard.edu/abs/2012MNRAS.425.2931O} {425, 2931}

\bibitem[\protect\citeauthoryear{{Owen} \& {Kollmeier}}{{Owen} \& {Kollmeier}}{2017}]{Owen2017}
{Owen} J.~E.,  {Kollmeier} J.~A.,  2017, \mn@doi [\mnras] {10.1093/mnras/stx302}, \href {http://adsabs.harvard.edu/abs/2017MNRAS.467.3379O} {467, 3379}

\bibitem[\protect\citeauthoryear{{Owen} \& {Schlichting}}{{Owen} \& {Schlichting}}{2024}]{Owen2024}
{Owen} J.~E.,  {Schlichting} H.~E.,  2024, \mn@doi [\mnras] {10.1093/mnras/stad3972}, \href {https://ui.adsabs.harvard.edu/abs/2024MNRAS.528.1615O} {528, 1615}

\bibitem[\protect\citeauthoryear{{Owen} \& {Wu}}{{Owen} \& {Wu}}{2013}]{Owen2013}
{Owen} J.~E.,  {Wu} Y.,  2013, \mn@doi [\apj] {10.1088/0004-637X/775/2/105}, \href {http://adsabs.harvard.edu/abs/2013ApJ...775..105O} {775, 105}

\bibitem[\protect\citeauthoryear{{Pecaut} \& {Mamajek}}{{Pecaut} \& {Mamajek}}{2013}]{Pecaut2013}
{Pecaut} M.~J.,  {Mamajek} E.~E.,  2013, \mn@doi [\apjs] {10.1088/0067-0049/208/1/9}, \href {http://adsabs.harvard.edu/abs/2013ApJS..208....9P} {208, 9}

\bibitem[\protect\citeauthoryear{{Petigura}}{{Petigura}}{2020}]{Petigura2020}
{Petigura} E.~A.,  2020, \mn@doi [\aj] {10.3847/1538-3881/ab9fff}, \href {https://ui.adsabs.harvard.edu/abs/2020AJ....160...89P} {160, 89}

\bibitem[\protect\citeauthoryear{{Petigura} et~al.,}{{Petigura} et~al.}{2022}]{Petigura2022}
{Petigura} E.~A.,  et~al., 2022, \mn@doi [\aj] {10.3847/1538-3881/ac51e3}, \href {https://ui.adsabs.harvard.edu/abs/2022AJ....163..179P} {163, 179}

\bibitem[\protect\citeauthoryear{{Quintana} et~al.,}{{Quintana} et~al.}{2014}]{Quintana2014}
{Quintana} E.~V.,  et~al., 2014, \mn@doi [Science] {10.1126/science.1249403}, \href {https://ui.adsabs.harvard.edu/abs/2014Sci...344..277Q} {344, 277}

\bibitem[\protect\citeauthoryear{{Ravinet}, {Lagarde}, {Reyl{\'e}}, {Amard}  \& {Van Grootel}}{{Ravinet} et~al.}{2022}]{Ravinet2021}
{Ravinet} T.,  {Lagarde} N.,  {Reyl{\'e}} C.,  {Amard} L.,   {Van Grootel} V.,  2022, in Cambridge Workshop on Cool Stars, Stellar Systems, and the Sun. Cambridge Workshop on Cool Stars, Stellar Systems, and the Sun.
p.~166, \mn@doi{10.5281/zenodo.7586227}

\bibitem[\protect\citeauthoryear{{Reinhold} \& {Hekker}}{{Reinhold} \& {Hekker}}{2020}]{Reinhold2020}
{Reinhold} T.,  {Hekker} S.,  2020, \mn@doi [\aap] {10.1051/0004-6361/201936887}, \href {https://ui.adsabs.harvard.edu/abs/2020A&A...635A..43R} {635, A43}

\bibitem[\protect\citeauthoryear{{Rogers}}{{Rogers}}{2015}]{Rogers2015}
{Rogers} L.~A.,  2015, \mn@doi [\apj] {10.1088/0004-637X/801/1/41}, \href {http://adsabs.harvard.edu/abs/2015ApJ...801...41R} {801, 41}

\bibitem[\protect\citeauthoryear{{Rogers} \& {Seager}}{{Rogers} \& {Seager}}{2010}]{Rogers2010}
{Rogers} L.~A.,  {Seager} S.,  2010, \mn@doi [\apj] {10.1088/0004-637X/712/2/974}, \href {https://ui.adsabs.harvard.edu/abs/2010ApJ...712..974R} {712, 974}

\bibitem[\protect\citeauthoryear{{Sandoval}, {Contardo}  \& {David}}{{Sandoval} et~al.}{2021}]{Sandoval2021}
{Sandoval} A.,  {Contardo} G.,   {David} T.~J.,  2021, \mn@doi [\apj] {10.3847/1538-4357/abea9e}, \href {https://ui.adsabs.harvard.edu/abs/2021ApJ...911..117S} {911, 117}

\bibitem[\protect\citeauthoryear{{Santos}, {Garc{\'\i}a}, {Mathur}, {Bugnet}, {van Saders}, {Metcalfe}, {Simonian}  \& {Pinsonneault}}{{Santos} et~al.}{2019}]{Santos2019}
{Santos} A.~R.~G.,  {Garc{\'\i}a} R.~A.,  {Mathur} S.,  {Bugnet} L.,  {van Saders} J.~L.,  {Metcalfe} T.~S.,  {Simonian} G.~V.~A.,   {Pinsonneault} M.~H.,  2019, \mn@doi [\apjs] {10.3847/1538-4365/ab3b56}, \href {https://ui.adsabs.harvard.edu/abs/2019ApJS..244...21S} {244, 21}

\bibitem[\protect\citeauthoryear{{Schaefer}, {Wordsworth}, {Berta-Thompson}  \& {Sasselov}}{{Schaefer} et~al.}{2016}]{Schaefer2016}
{Schaefer} L.,  {Wordsworth} R.~D.,  {Berta-Thompson} Z.,   {Sasselov} D.,  2016, \mn@doi [\apj] {10.3847/0004-637X/829/2/63}, \href {https://ui.adsabs.harvard.edu/abs/2016ApJ...829...63S} {829, 63}

\bibitem[\protect\citeauthoryear{{Schlafly} \& {Finkbeiner}}{{Schlafly} \& {Finkbeiner}}{2011}]{Schlafly2011}
{Schlafly} E.~F.,  {Finkbeiner} D.~P.,  2011, \mn@doi [\apj] {10.1088/0004-637X/737/2/103}, \href {https://ui.adsabs.harvard.edu/abs/2011ApJ...737..103S} {737, 103}

\bibitem[\protect\citeauthoryear{{Scott}}{{Scott}}{1992}]{Scott1992}
{Scott} D.~W.,  1992, {Multivariate Density Estimation}.
{John Wiley \& Sons}

\bibitem[\protect\citeauthoryear{{Silverberg} et~al.,}{{Silverberg} et~al.}{2020}]{Silverberg2020}
{Silverberg} S.~M.,  et~al., 2020, \mn@doi [\apj] {10.3847/1538-4357/ab68e6}, \href {https://ui.adsabs.harvard.edu/abs/2020ApJ...890..106S} {890, 106}

\bibitem[\protect\citeauthoryear{{Sinukoff} et~al.,}{{Sinukoff} et~al.}{2017}]{Sinukoff2017}
{Sinukoff} E.,  et~al., 2017, \mn@doi [\aj] {10.3847/1538-3881/aa725f}, \href {http://adsabs.harvard.edu/abs/2017AJ....153..271S} {153, 271}

\bibitem[\protect\citeauthoryear{{Skrutskie} et~al.,}{{Skrutskie} et~al.}{2006}]{Skrutskie2006}
{Skrutskie} M.~F.,  et~al., 2006, \mn@doi [\aj] {10.1086/498708}, \href {http://adsabs.harvard.edu/abs/2006AJ....131.1163S} {131, 1163}

\bibitem[\protect\citeauthoryear{{Skumanich}}{{Skumanich}}{1972}]{Skumanich1972}
{Skumanich} A.,  1972, \mn@doi [\apj] {10.1086/151310}, \href {https://ui.adsabs.harvard.edu/abs/1972ApJ...171..565S} {171, 565}

\bibitem[\protect\citeauthoryear{{Smart} et~al.,}{{Smart} et~al.}{2021}]{Smart2021}
{Smart} R.,  et~al., 2021, in The 20.5th Cambridge Workshop on Cool Stars, Stellar Systems, and the Sun (CS20.5). Cambridge Workshop on Cool Stars, Stellar Systems, and the Sun.
p.~81, \mn@doi{10.5281/zenodo.4562738}

\bibitem[\protect\citeauthoryear{{Stauffer}, {Rebull}, {Cody}, {Hillenbrand}, {Pinsonneault}, {Barrado}, {Bouvier}  \& {David}}{{Stauffer} et~al.}{2018}]{Stauffer2018b}
{Stauffer} J.,  {Rebull} L.~M.,  {Cody} A.~M.,  {Hillenbrand} L.~A.,  {Pinsonneault} M.,  {Barrado} D.,  {Bouvier} J.,   {David} T.,  2018, \mn@doi [\aj] {10.3847/1538-3881/aae9ec}, \href {https://ui.adsabs.harvard.edu/abs/2018AJ....156..275S} {156, 275}

\bibitem[\protect\citeauthoryear{{Sullivan} et~al.,}{{Sullivan} et~al.}{2023}]{Sullivan2023}
{Sullivan} K.,  et~al., 2023, \mn@doi [\aj] {10.3847/1538-3881/acbdf9}, \href {https://ui.adsabs.harvard.edu/abs/2023AJ....165..177S} {165, 177}

\bibitem[\protect\citeauthoryear{{Swift}, {Johnson}, {Morton}, {Crepp}, {Montet}, {Fabrycky}  \& {Muirhead}}{{Swift} et~al.}{2013}]{Swift2013}
{Swift} J.,  {Johnson} J.~A.,  {Morton} T.,  {Crepp} J.~R.,  {Montet} B.,  {Fabrycky} D.~C.,   {Muirhead} P.,  2013, in American Astronomical Society Meeting Abstracts \#221. p. 407.04

\bibitem[\protect\citeauthoryear{{Szab{\'o}} \& {Kiss}}{{Szab{\'o}} \& {Kiss}}{2011}]{Szabo2011}
{Szab{\'o}} G.~M.,  {Kiss} L.~L.,  2011, \mn@doi [\apjl] {10.1088/2041-8205/727/2/L44}, \href {https://ui.adsabs.harvard.edu/abs/2011ApJ...727L..44S} {727, L44}

\bibitem[\protect\citeauthoryear{{Tabone} et~al.,}{{Tabone} et~al.}{2023}]{Tabone2023}
{Tabone} B.,  et~al., 2023, \mn@doi [Nature Astronomy] {10.1038/s41550-023-01965-3}, \href {https://ui.adsabs.harvard.edu/abs/2023NatAs...7..805T} {7, 805}

\bibitem[\protect\citeauthoryear{{Teixeira} \& {Ballard}}{{Teixeira} \& {Ballard}}{2023}]{Teixeira2023}
{Teixeira} K.,  {Ballard} S.,  2023, \mn@doi [\apj] {10.3847/1538-4357/acdc20}, \href {https://ui.adsabs.harvard.edu/abs/2023ApJ...953...50T} {953, 50}

\bibitem[\protect\citeauthoryear{{Thompson} et~al.,}{{Thompson} et~al.}{2018}]{Thompson2018}
{Thompson} S.~E.,  et~al., 2018, \mn@doi [\apjs] {10.3847/1538-4365/aab4f9}, \href {http://adsabs.harvard.edu/abs/2018ApJS..235...38T} {235, 38}

\bibitem[\protect\citeauthoryear{{Tonry} et~al.,}{{Tonry} et~al.}{2012}]{Tonry2012}
{Tonry} J.~L.,  et~al., 2012, \mn@doi [\apj] {10.1088/0004-637X/750/2/99}, \href {https://ui.adsabs.harvard.edu/abs/2012ApJ...750...99T} {750, 99}

\bibitem[\protect\citeauthoryear{{Turbet}, {Ehrenreich}, {Lovis}, {Bolmont}  \& {Fauchez}}{{Turbet} et~al.}{2019}]{Turbet2019}
{Turbet} M.,  {Ehrenreich} D.,  {Lovis} C.,  {Bolmont} E.,   {Fauchez} T.,  2019, \mn@doi [\aap] {10.1051/0004-6361/201935585}, \href {https://ui.adsabs.harvard.edu/abs/2019A&A...628A..12T} {628, A12}

\bibitem[\protect\citeauthoryear{{Turbet}, {Bolmont}, {Ehrenreich}, {Gratier}, {Leconte}, {Selsis}, {Hara}  \& {Lovis}}{{Turbet} et~al.}{2020}]{Turbet2020}
{Turbet} M.,  {Bolmont} E.,  {Ehrenreich} D.,  {Gratier} P.,  {Leconte} J.,  {Selsis} F.,  {Hara} N.,   {Lovis} C.,  2020, \mn@doi [\aap] {10.1051/0004-6361/201937151}, \href {https://ui.adsabs.harvard.edu/abs/2020A&A...638A..41T} {638, A41}

\bibitem[\protect\citeauthoryear{{Valizadegan}, {Martinho}, {Jenkins}, {Caldwell}, {Twicken}  \& {Bryson}}{{Valizadegan} et~al.}{2023}]{Valizadegan2023}
{Valizadegan} H.,  {Martinho} M. J.~S.,  {Jenkins} J.~M.,  {Caldwell} D.~A.,  {Twicken} J.~D.,   {Bryson} S.~T.,  2023, \mn@doi [\aj] {10.3847/1538-3881/acd344}, \href {https://ui.adsabs.harvard.edu/abs/2023AJ....166...28V} {166, 28}

\bibitem[\protect\citeauthoryear{{Van Eylen} et~al.,}{{Van Eylen} et~al.}{2019}]{VanEylen2019}
{Van Eylen} V.,  et~al., 2019, \mn@doi [\aj] {10.3847/1538-3881/aaf22f}, \href {https://ui.adsabs.harvard.edu/abs/2019AJ....157...61V} {157, 61}

\bibitem[\protect\citeauthoryear{{Van Eylen} et~al.,}{{Van Eylen} et~al.}{2021}]{VanEylen2021}
{Van Eylen} V.,  et~al., 2021, \mn@doi [\mnras] {10.1093/mnras/stab2143}, \href {https://ui.adsabs.harvard.edu/abs/2021MNRAS.507.2154V} {507, 2154}

\bibitem[\protect\citeauthoryear{{Vazan}, {Ormel}, {Noack}  \& {Dominik}}{{Vazan} et~al.}{2018}]{Vazan2018b}
{Vazan} A.,  {Ormel} C.~W.,  {Noack} L.,   {Dominik} C.,  2018, \mn@doi [\apj] {10.3847/1538-4357/aaef33}, \href {https://ui.adsabs.harvard.edu/abs/2018ApJ...869..163V} {869, 163}

\bibitem[\protect\citeauthoryear{{Vivien}, {Aguichine}, {Mousis}, {Deleuil}  \& {Marcq}}{{Vivien} et~al.}{2022}]{Vivien2022}
{Vivien} H.~G.,  {Aguichine} A.,  {Mousis} O.,  {Deleuil} M.,   {Marcq} E.,  2022, \mn@doi [\apj] {10.3847/1538-4357/ac66e2}, \href {https://ui.adsabs.harvard.edu/abs/2022ApJ...931..143V} {931, 143}

\bibitem[\protect\citeauthoryear{{Weiss} \& {Marcy}}{{Weiss} \& {Marcy}}{2014}]{Weiss2014}
{Weiss} L.~M.,  {Marcy} G.~W.,  2014, \mn@doi [\apjl] {10.1088/2041-8205/783/1/L6}, \href {http://adsabs.harvard.edu/abs/2014ApJ...783L...6W} {783, L6}

\bibitem[\protect\citeauthoryear{{Weiss} et~al.,}{{Weiss} et~al.}{2018}]{Weiss2018}
{Weiss} L.~M.,  et~al., 2018, \mn@doi [\aj] {10.3847/1538-3881/aa9ff6}, \href {https://ui.adsabs.harvard.edu/abs/2018AJ....155...48W} {155, 48}

\bibitem[\protect\citeauthoryear{{Yao} et~al.,}{{Yao} et~al.}{2021}]{Yao2021}
{Yao} X.,  et~al., 2021, \mn@doi [\aj] {10.3847/1538-3881/abdb30}, \href {https://ui.adsabs.harvard.edu/abs/2021AJ....161..124Y} {161, 124}

\bibitem[\protect\citeauthoryear{{Yoshida}, {Terada}, {Ikoma}  \& {Kuramoto}}{{Yoshida} et~al.}{2022}]{Yoshida2022}
{Yoshida} T.,  {Terada} N.,  {Ikoma} M.,   {Kuramoto} K.,  2022, \mn@doi [\apj] {10.3847/1538-4357/ac7be7}, \href {https://ui.adsabs.harvard.edu/abs/2022ApJ...934..137Y} {934, 137}

\bibitem[\protect\citeauthoryear{{Yuan}, {Liu}  \& {Xiang}}{{Yuan} et~al.}{2013}]{Yuan2013}
{Yuan} H.~B.,  {Liu} X.~W.,   {Xiang} M.~S.,  2013, \mn@doi [\mnras] {10.1093/mnras/stt039}, \href {http://adsabs.harvard.edu/abs/2013MNRAS.430.2188Y} {430, 2188}

\bibitem[\protect\citeauthoryear{{Zhu} \& {Dong}}{{Zhu} \& {Dong}}{2021}]{ZhuW2021}
{Zhu} W.,  {Dong} S.,  2021, \mn@doi [\araa] {10.1146/annurev-astro-112420-020055}, \href {https://ui.adsabs.harvard.edu/abs/2021ARA&A..59..291Z} {59, 291}

\bibitem[\protect\citeauthoryear{{Zieba} et~al.,}{{Zieba} et~al.}{2023}]{Zieba2023}
{Zieba} S.,  et~al., 2023, \mn@doi [\nat] {10.1038/s41586-023-06232-z}, \href {https://ui.adsabs.harvard.edu/abs/2023Natur.620..746Z} {620, 746}

\bibitem[\protect\citeauthoryear{{Ziegler} et~al.,}{{Ziegler} et~al.}{2018a}]{Ziegler2018}
{Ziegler} C.,  et~al., 2018a, \mn@doi [\aj] {10.3847/1538-3881/aace59}, \href {https://ui.adsabs.harvard.edu/abs/2018AJ....156...83Z} {156, 83}

\bibitem[\protect\citeauthoryear{{Ziegler} et~al.,}{{Ziegler} et~al.}{2018b}]{Ziegler2018b}
{Ziegler} C.,  et~al., 2018b, \mn@doi [\aj] {10.3847/1538-3881/aad80a}, \href {https://ui.adsabs.harvard.edu/abs/2018AJ....156..259Z} {156, 259}

\makeatother
\end{thebibliography}
